\documentclass[12pt,draft]{article}

\usepackage{amsthm,amssymb,amsmath,amscd,amstext,mathrsfs}
\usepackage{latexsym, mathptmx, xypic}

\newtheorem{definition}{Definition}[section]
\newtheorem{theorem}{Theorem}[section]
\newtheorem{lemma}{Lemma}[section]
\newtheorem*{remark}{{\it Remark}}

\newcommand{\nc}{\newcommand}

\nc{\stack}[2]{{\begin{array}{c}
\scriptstyle #1 \\ \scriptstyle #2 \end{array}} }

\nc{\C}{{\mathbb C}}  
\nc{\R}{{\mathbb R}}  
\nc{\HH}{{\mathbb H}} 
\nc{\Z}{{\mathbb Z}}  
\nc{\N}{{\mathbb N}}
\nc{\dd}{{\rm d}}
\nc{\DD}{{\rm D}}
\nc{\ii}{{\bf i}}
\nc{\ca}{{\mathscr A}}
\nc{\cf}{{\mathscr B}}
\nc{\cc}{{\mathscr C}}
\nc{\cd}{{\mathscr D}}
\nc{\ce}{{\mathscr E}}
\nc{\ch}{{\mathscr H}}
\nc{\cm}{{\mathscr M}}
\nc{\co}{{\mathscr O}}
\nc{\cu}{{\mathscr U}}
\nc{\cv}{{\mathscr V}}
\nc{\cz}{{\mathscr Z}}

\nc{\cb}{{\mathfrak B}}
\nc{\fu}{{\mathfrak A}}

\begin{document} 

\title{Gravity as a four dimensional algebraic quantum\\
                          field theory}

\author{G\'abor Etesi\\
\small{{\it Department of Geometry, Mathematical Institute, Faculty of
Science,}}\\
\small{{\it Budapest University of Technology and Economics,}}\\
\small{{\it Egry J. u. 1, H \'ep., H-1111 Budapest, Hungary}}
\footnote{e-mail: {\tt etesi@math.bme.hu}}}

\maketitle

\pagestyle{myheadings}
\markright{G. Etesi: Gravity as an algebraic quantum field theory}

\thispagestyle{empty}

\begin{abstract} Based on a family of indefinite unitary representations of 
the diffeomorphism group of an oriented smooth $4$-manifold, a manifestly
covariant $4$ dimensional and non-perturbative algebraic quantum field 
theory formulation of gravity is exhibited. More precisely 
among the bounded linear operators acting on these representation spaces we 
identify algebraic curvature tensors hence a net of local quantum 
observables can be constructed from $C^*$-algebras generated by local 
curvature tensors and vector fields. This algebraic quantum field theory 
is extracted from structures provided by an oriented smooth $4$-manifold 
only hence possesses a diffeomorphism symmetry. In this way classical general 
relativity exactly in $4$ dimensions naturally embeds into a quantum framework.

Several Hilbert space representations of the theory are found. First a 
``tautological representation'' of the limiting global $C^*$-algebra 
is constructed allowing to associate to any oriented smooth $4$-manifold a 
von Neumann algebra in a canonical fashion. Secondly, 
influenced by the Dougan--Mason approach to gravitational quasilocal 
energy-momentum, we construct certain representations what we call 
``positive mass representations'' with unbroken diffeomorphism symmetry. 
Thirdly, we also obtain ``classical representaions'' with spontaneously 
broken diffeomorphism symmetry corresponding to the classical limit of the 
theory which turns out to be general relativity.

Finally we observe that the whole family of ``positive mass 
representations'' comprise a $2$ dimensional conformal field 
theory in the sense of G. Segal. 

\end{abstract}

\centerline{AMS Classification: Primary: 83C45, 81T05, Secondary: 57N13}
\centerline{Keywords: {\it General relativity; Algebraic quantum 
field theory; Four dimensions}}


\section{Introduction}
\label{one}


The outstanding problem of modern theoretical physics is how to unify 
the obviously successful and mathematically consistent 
{\it theory of general relativity} with the obviously successful but yet 
mathematically problematic {\it relativistic quantum field theory}. It 
has been generally believed that these two fundamental pillars of modern 
theoretical physics conflict each other not only in the mathematical 
tools they use but even at a deep foundational level \cite{hed}: classical 
concepts of general relativity such as the space-time event, the light 
cone or the event horizon of a black hole are too ``sharp'' objects from a 
quantum theoretic viewpoint meanwhile relativistic quantum field theory is not 
background independent from the aspect of general relativity. We do 
not attempt here to survey the vast physical and even mathematical and 
philosophical literature created by the unification problem; we just mention 
that nowadays the two leading candidates expected to be capable for a 
sort of unification are string theory and loop quantum gravity. But surely 
there is still a long way ahead; nevertheless we have the conviction that one 
day the language of classical general relativity will sound familiar to 
quantum theorists and vice versa i.e., conceptual bridges must exist 
connecting the two theories. 

In this note an effort has been made to embed classical general relativity 
into a quantum framework. This quantum framework is 
{\it algebraic quantum field theory} formulated by Haag--Kastler and others 
during the past decades, cf. \cite{haa}. Recently this language also appears 
to be suitable for formulating quantum field theory on curved 
space-time \cite{bru-fre-ver, hol-wal} or even quantum gravity 
\cite{bru-fre-rej}.

In more detail we will do something very simple here. Namely using 
structures provided by an oriented smooth $4$-manifold $M$ only, our overall 
guiding principle will be seeking unitary representations of the corresponding 
orientation-preserving diffeomorphism group ${\rm Diff}^+(M)$. There is a 
unique such representation via pullback on the incomplete space of 
sections of $\wedge^2M\otimes_\R\C$. However the natural scalar product 
on this space---namely the one given by integration of the wedge product of 
two $2$-forms---is indefinite hence cannot be used to complete the space of 
smooth $2$-forms into a Hilbert space. Rather in struggling with the 
completion problem one comes up with a family of Hilbert spaces with a 
common non-degenerate indefinite Hermitian scalar product on them. 
The bare Hilbert spaces---i.e., not considered as 
${\rm Diff}^+(M)$-modules---admit decompositions 
$\ch^+(M)\oplus\ch^-(M)$ into maximal definite orthogonal Hilbert 
subspaces $\ch^\pm (M)$ with respect to the indefinite scalar product. One can 
use this family of Hilbert spaces to discover an interesting $C^*$-algebra by 
exploring their spaces of bounded linear operators. It indeed comes as a 
surprise (at least to the author) that precisely in $4$ dimensions among these 
operators one can recognize curvature tensors! This is because of the 
well-known fact that the curvature tensor $R_g$ of a pseudo-Riemannian 
$4$-manifold $(M,g)$ can be viewed as a section of 
${\rm End}(\wedge^2M\otimes_\R\C )$ i.e., gives rise to a linear operator 
acting on any $\ch^+(M)\oplus\ch^-(M)$. This permits to construct a net 
$\{ U\mapsto\fu (U)\}_{U\subseteqq M}$ whose local $C^*$-algebras are 
generated by bundle endomorphisms and Lie 
derivatives. These local algebras are generalizations of the CCR algebra. 
The construction satisfies the naturally generalized Haag--Kastler 
axioms \cite[pp. 105-107]{haa} leading to an algebraic quantum field theory in 
which Poincar\'e symmetry is replaced with full diffeomorphism symmetry (if 
the diffeomorphism group is regarded as the physical symmetry group of 
general relativity and not its gauge group). As a 
result classical general relativity effortlessly embeds into a quantum 
framework if one interprets classical curvature tensors as quantum observables. 
The appearence of the curvature tensor as a local quantum observable is 
reasonable even from the physical viewpoint: in {\it local} gravitational 
physics the metric tensor has no direct physical meaning only its curvature 
can cause local physical effects such as tidal forces. Moreover if one wishes, 
at least in principle, the metric i.e., the geometry locally can be 
reconstructed from its curvature (see e.g. \cite{det,gas,hal,kow} and the 
references therein).

We also exhibit several Hilbert space representations of the theory carrying 
unitary representations of the diffeomorphism group. The 
first one is a ``tautological representation'' of the global algebra on 
itself allowing us to attach to $M$ a von Neumann algebra ${\mathfrak R}(M)$. 
The other ones deal with physics. A meaningful quantum field 
theory must exhibit stability i.e., ``positive mass representations'' of 
its local observables in the sense of Wigner. In our case this directly 
leads to the long-standing problem of gravitational mass \cite{sza}. It is 
quite interesting that the Gelfand--Naimark--Segal construction in the 
theory of $C^*$-algebras and quasilocal energy-momentum constructions 
\cite{sza} in general relativity naturally meet up here because immersed 
surfaces in $M$ provide us with {\it both} $C^*$-algebra representations 
and Dougan--Mason-like quasilocal quantities \cite{dou-mas}. More 
precisely our quasilocal energy-momenta and masses stem from quasilocal 
translations along immersed surfaces in $M$ with a choice of a complex 
structure on them. However the whole construction is expected to be 
independent of this choice leading to the by-now classical observation of 
Witten \cite{wit} that in fact one has to deal with a conformal field 
theory on these surfaces. We identify this theory: its spaces of conformal 
blocks are the Clifford algebras generated by finite energy meromorphic 
sections of certain unitary holomorphic vector bundles on punctured 
Riemannian surfaces. Apart from these quantum representations, ``classical 
representations'' corresponding to the classical limit of the theory also 
exist. Indeed, unlike in the previous two cases, in these representations the 
diffeomorphism symmetry spontaneously breaks down to a finite dimensional 
Lie subgroup provided by the isometry group of an emergent metric $g$ on $M$; 
hence a causal structure can be constructed on $M$ as well. Therefore 
classical general relativity is recovered again at the representation 
theoretic level.  The emergent metric distinguishes a canonically split 
Hilbert space $\ch^+(M)\oplus\ch^-(M)$ provided by metric 
(anti)self-duality leading to a splitting of the Hilbert space of the 
corresponding ``classical representation'', too. The natural quantum 
observable provided by the curvature $R_g$ of the metric in this 
representation obeys the splitting if and only if $g$ is a vacuum metric.

However our algebraic quantum field theory itself lacks any causal 
structure in general as an unavoidable consequence of its vast 
diffeomorphism symmetry.\footnote{This is in accordance with recent 
speculations on Lorentz symmetry violations for instance in extreme high 
energy cosmic processes, for a review cf. e.g. \cite{car-mig}.} The causal 
future $J^+(p)\subset M$ of an event $p\in M$ in space-time is by 
definition the union of all future-inextendible worldlines of particles 
departing from $p$ and moving forward in time locally not exceeding the 
speed of light. The causal past $J^-(p)$ is defined similarly. The 
collection of these subsets of space-time generates a special topology on 
$M$ in the strict mathematical sense. The Lorentzian metric is a 
mathematical fusion of the geometry of $M$ identified with a Riemannian 
metric {\it and} the causal structure of $M$ identified with this 
topology. But from this operational description of causality it is clear 
that the construction of a causal structure refers to not only gravity but 
other entities of physical reality as well which are moreover quite 
classical: pointlike particles, electromagnetic waves, time, etc. However 
they cannot appear for instance in a vacuum space-time considered in the 
strict sense. Very strictly speaking even the interpretation of a 
space-time point as a ``physical event'' fails in an {\it empty} 
space-time. Therefore we are convinced that causality cannot be a 
fundamental ingredient of a classical hence even of a quantum description 
of pure gravity if it is a diffeomorphism-invariant quantum field theory. 
As a technical consequence we will prefer to use Riemannian metrics in 
this note (although emphasize that mathematically all conclusions hold for 
Lorentzian metrics as well). To summarize: from our standpoint causality 
is an emergent statistical phenomenon created by the highly complex 
interaction of gravity and matter. Consequently in order to recover it 
first we should be able to break down the diffeomorphism symmetry and 
distinguish pure gravity from matter.

This note is organized as follows. In Sect. \ref{two} we construct 
natural indefinite unitary representations of orientation-preserving 
diffeomorphisms of an oriented $4$-manifold. Then we extract a unique 
$C^*$-algebra out of these representation spaces. We identify its 
``classical part'' with Einstein manifolds. In Sect. \ref{three} we introduce 
an algebraic quantum field theory and in Theorems \ref{diff-top-reprezentacio} 
and \ref{pozitiv-tomegu-reprezentacio} and \ref{klasszikus-reprezentacio} we 
construct certain representations of its algebras of local observables what we 
call ``a tautological representation'', ``positive mass representations'' and 
``classical representations'' respectively. In Sect. \ref{four} we bunch 
the positive mass representations together into a conformal field theory. 
\vspace{0.1in}

\noindent{\bf Acknowledgement.} The author is grateful to M.J. Dupr\'e, I. 
Ojima, L.B. Szabados and P. Vrana for the stimulating discussions and to the 
Alfr\'ed R\'enyi Institute of Mathematics for their hospitality. 
This work was supported by OTKA grant No. NK81203 (Hungary).


\section{The $C^*$-algebra of an oriented smooth $4$-manifold}
\label{two}


Let $M$ be a connected orientable smooth $4$-manifold, possibly non-closed  
(i.e., it can be non-compact and-or with non-empty boundary). Fix an 
orientation on $M$. Given only these data at our disposal it is already 
meaningful to talk about the group of its orientation-preserving 
diffeomorphisms ${\rm Diff}^+(M)$. Our overall guiding principle simply 
will be a search for unitary representations of ${\rm Diff}^+(M)$. 
A bunch of representations arise in a geometric way as follows. 
Consider $T^{(r,s)}M\otimes_\R\C$, the bundle of complexified $(r,s)$-type 
tensors with the associated vector spaces 
$C^\infty_c(M; T^{(r,s)}M\otimes_\R\C )$ of their compactly supported 
smooth complexified sections. Then the group ${\rm Diff}^+(M)$ acts 
from the left via pushforward on $C^\infty_c(M; T^{(r,0)}M\otimes_\R\C )$ 
for all $r\in\N$ while from the right via pullback on $C^\infty_c(M; 
T^{(0,s)}M\otimes_\R\C )$ for all $s\in\N$. However these representations are 
typically not unitary because the underlying vector spaces do not carry extra 
structures in a natural way. 

The only exception is the $2^{{\rm nd}}$ exterior power 
$\wedge^2M\subset T^{(0,2)}M$ of the cotangent bundle with the 
corresponding space of sections $C^\infty_c(M;\wedge^2M\otimes_\R\C )
=:\Omega^2_c(M;\C )$, the space of complexified smooth $2$-forms with 
compact support. Indeed, this vector space has a natural non-degenerate 
Hermite scalar product 
$\langle\:\cdot\:,\:\cdot\:\rangle_{L^2(M)}:\:\:\:\:\:\Omega^2_c(M;\C )\times 
\Omega^2_c(M;\C )\rightarrow\C$ given by integration on oriented smooth 
manifolds; more precisely for $\alpha, \beta\in\Omega^2_c(M;\C )$ put
\begin{equation}
\langle\alpha ,\beta\rangle_{L^2(M)}:=\int\limits_M\overline{\alpha}\wedge\beta
\label{integralas}
\end{equation}
(complex conjugate-linear in its first variable). Note however that this 
scalar product is {\it indefinite}: an unavoidable fact which plays a key role 
in our considerations ahead. Consequently this scalar product cannot be used 
to complete $\Omega^2_c(M;\C )$ into a Hilbert space. Instead with respect to 
(\ref{integralas}) there is a non-unique direct sum decomposition 
\[\Omega^2_c(M;\C )=\Omega^+ _c(M;\C )\oplus \Omega^-_c(M;\C )\]
with the property that they are maximal definite orthogonal subspaces i.e., 
$\pm\langle\:\cdot\:,\:\cdot\:\rangle_{L^2(M)}\vert_{\Omega^\pm _c(M;\C )}
\:\:\::\:\:\: \Omega^\pm _c(M;\C )\times \Omega^\pm _c(M;\C )\rightarrow\C$ 
are both positive definite moreover $\Omega^+ _c(M;\C )
\perp_{L^2(M)}\Omega^-_c(M;\C )$. Therefore these restricted scalar products 
can be used to complete $\Omega^\pm _c(M;\C )$ into separable Hilbert spaces 
$\ch^\pm (M)$ respectively. That is, starting with an $M$ we can make 
$\Omega^2_c(M;\C )$ complete only in non-canonical ways as follows. The 
possible completions form a family and any member of this family consists 
of a particular direct sum Hilbert space $\ch^+(M)\oplus\ch^-(M)$ (with its 
particular non-degenerate positive definite scalar product 
$(\alpha ,\beta )_{L^2(M)}:=
\langle\alpha^+,\beta^+\rangle_{L^2(M)}-\langle\alpha^-,
\beta^-\rangle_{L^2(M)}$) and a common indefinite scalar product
\begin{equation}
\langle\:\cdot\:,\:\cdot\:\rangle_{L^2(M)}:\ch^+(M)\oplus\ch^-(M) \times 
\ch^+(M)\oplus\ch^-(M)\longrightarrow\C
\label{skalarszorzas}
\end{equation}
induced by (\ref{integralas}) such that:
\[\left\{\begin{array}{ll}
\ch^+(M)\perp_{L^2(M)}\ch^-(M),\\
\mbox{$\langle\:\cdot\:,\:\cdot\:\rangle_{L^2(M)}\vert_{\ch^\pm (M)}:
\ch^\pm (M)\times\ch^\pm (M)\longrightarrow\C$ are positive or negative 
definite, respectively.}
\end{array}\right.\] 

\noindent Moreover any $(\ch^+(M)\oplus\ch^-(M), 
\langle\:\cdot\:,\:\cdot\:\rangle_{L^2(M)})$ carries a representation of 
${\rm Diff}^+(M)$ from the right given by the unique continuous 
extension of the pullback of $2$-forms: 
$\omega\mapsto f^*\omega$ for $\omega\in\Omega^2_c(M;\C )$ and 
$f\in{\rm Diff}^+(M)$. It is easy to check that these operators are unitary 
with respect to (\ref{skalarszorzas}) and operators  
corresponding to compactly supported diffeomorphisms are also 
bounded with respect to the operator norm induced by the particular Hilbert 
space norm on $\ch^+(M)\oplus\ch^-(M)$. Note that {\it a priori}  
representations on different completions are {\it not} unitary equivalent.

These representations have the following immediate properties:

\begin{lemma} Consider the indefinite unitary reprsentation of 
${\rm Diff}^+(M)$ from the right on any particular 
$(\ch^+(M)\oplus\ch^-(M), \langle\:\cdot\:,\:\cdot\:\rangle_{L^2(M)})$ 
constructed above. 
\begin{itemize}

\item[(i)] A vector $v\in\ch^+(M)\oplus\ch^-(M)$ satisfies 
$f^*v= v$ for all $f\in{\rm Diff}^+(M)$ if and only if $v=0$ (``no vacuum''); 

\item[(ii)] The closed subspaces $\cf (M)\subseteqq\cz 
(M)\subset\ch^+(M)\oplus\ch^-(M)$ generated by exact or closed $2$-forms 
respectively are invariant under the action of ${\rm Diff}^+(M)$. 
\end{itemize}
\label{vakuum}
\end{lemma}

\noindent{\it Proof.} (i) Assume that there exists an element 
$0\not=v\in\ch^+(M)\oplus\ch^-(M)$ stabilized by the whole 
${\rm Diff}^+(M)$. Consider a $1$-parameter subgroup $\{f_t\}_{t\in\R}\in
{\rm Diff}^+(M)$ such that $f_0={\rm Id}_M$ and let $X$ be the vector field on 
$M$ generating this subgroup. Differentiating the equation $f^*_tv 
=v$ with respect to $t\in\R$ at $t=0$ we obtain $L_Xv=0$ 
(in the weak sense) where $L_X$ is the Lie derivative by $X$. Since 
an arbitrary compactly supported vector field generates a $1$-parameter 
subgroup of ${\rm Diff}^+(M)$ we obtain that in fact $v=0$, a contradiction.

(ii) The statement readily follows by naturality of exterior 
differentiation i.e., $\dd (f^*\varphi )=f^*\dd\varphi$ for all 
$f\in{\rm Diff}^+(M)$ and $\varphi\in\Omega^k_c(M;\C )$. $\Diamond$

\begin{remark}\rm 1. We succeeded to construct a family of faithful, 
reducible, indefinite unitary representations of the diffeomorphism group 
out of the structures provided only by an orientable smooth 
$4$-manifold.\footnote{In fact our construction so far works in any $4k$ 
($k=1,2,\dots$) dimensions if the diffeomorphism group acts on $2k$-forms. 
In $4k+2$ dimensions (\ref{integralas}) gives symplectic forms.} All of 
these representation spaces are split however such decompositions cannot 
hold as a ${\rm Diff}^+(M)$-module or in other words such decompositions 
break the diffeomorphism symmetry. The relevance of these splittings, as 
we will see shortly, is that the classical vacuum Einstein equation can be 
viewed as saying that there is a distinguished representation 
$\ch^+(M)\oplus\ch^-(M)$ on which the curvature is blockdiagonal i.e., 
respects the splitting. In general, starting only with an oriented smooth 
$4$-manifold $M$ without extra structure, there is no way to associate a 
canonical non-split Hilbert space to $M$.

2. From the mathematical viewpoint in many important cases we do not loose 
topological information if we replace $M$ with any representation. 
Indeed, restricting $\Omega^2_c(M;\C )$ to closed forms and dividing by 
the exact ones we can pass to compactly supported cohomology $H^2_c(M;\C )$; 
then if $M$ admits a finite good cover Poincar\'e duality works and gives 
$H^2_c(M;\C )\cong (H^2 (M;\C ))^*$. If we assume that $M$ is 
compact and simply connected then the singular cohomology $H^2(M;\Z )$ 
maps injectively into $H^2(M;\C )$ hence finally the scalar product 
(\ref{skalarszorzas}) descends to the topological intersection form 
\[q_M: H^2(M;\Z )\times H^2(M;\Z )\longrightarrow H^4(M;\Z )\cong\Z\] 
of the underlying topological $4$-manifold. However taking into account that by 
assumption $M$ has a smooth structure we can refer to Freedman's fundamental 
result \cite{fre} that $q_M$ uniquely determines the topology of $M$.

\end{remark}

\noindent Now we proceed further and observe that in spite of this 
{\it plethora} of diffeomorphism group representations one can attach a 
{\it unique} $C^*$-algebra to an oriented smooth $4$-manifold. However 
this $C^*$-algebra does not admit representations on the previous Hilbert 
spaces.

\begin{lemma} Let $\divideontimes$ be the adjoint operation on 
$\Omega^2_c(M;\C )$ for the indefinite scalar product (\ref{integralas}).
Consider the $\divideontimes$-closed space 
$V:=\left\{A\in{\rm End}(\Omega^2_c(M;\C ))
\:\vert\: r(A^\divideontimes A)<+\infty\right\}$ defined by the 
spectral radius  
\[r(B):=\sup\limits_{\lambda\in\C}
\left\{\mbox{$\vert\lambda\vert\:\:\left\vert\:B-
\lambda\cdot{\rm Id}_{\Omega^2_c(M;\C )}\right.$ {\rm is not invertible}}
\right\}\:\:\:.\]
Then $\sqrt{r}$ is a norm and the 
corresponding completion of $V$ renders $(V, \divideontimes )$ a unital 
$C^*$-algebra containing ${\rm Diff}^+(M)$. This $C^*$-algebra will be 
denoted by $\cb (M)$. 
\label{c*lemma}
\end{lemma}

\noindent {\it Proof.} Our strategy to prove the lemma is as follows. 
Obviously $(V,\divideontimes )$ is a $*$-algebra. Provided it can be 
equipped with a norm such that corresponding completion of 
$V$ improves $(V,\divideontimes )$ to a $C^*$-algebra then knowing the 
uniqueness of the $C^*$-algebra norm this sought norm $[[\:\cdot\:]]$ 
on all $A\in V$ must look like $[[A]]^2=[[A^\divideontimes A]]=
r(A^\divideontimes A)$. Therefore we want to see that the 
spectral radius gives a norm here. 

Take any splitting 
$\Omega^2_c(M;\C )=\Omega^+_c(M;\C )\oplus\Omega^-_c(M;\C )$ and the 
corresponding Hilbert space completion 
$\ch^+(M)\oplus\ch^-(M)\supset\Omega^2_c(M;\C )$. If 
$P^\pm :\ch^+(M)\oplus\ch^-(M)\rightarrow\ch^\pm (M)$ are the orthogonal 
projections then put $J:=P^+-P^-$ moreover let $\dagger$ denote the adjoint 
on $\ch^+(M)\oplus\ch^-(M)$. Then $J$ satisfies 
$A^\divideontimes =JA^\dagger J$ and $J^2={\rm Id}_{\ch^+(M)\oplus\ch^-(M)}$ 
therefore $A^\dagger =JA^\divideontimes J$ as well. Recall that the operator 
norm is  
\begin{equation}
\Vert B\Vert =\sup\limits_{v\not =0}\frac{\Vert Bv\Vert_{L^2(M)}}
{\Vert v\Vert_{L^2(M)}}
\label{opinorm}
\end{equation}
where $\Vert\:\cdot\:\Vert_{L^2(M)}$ comes from the positive
definite scalar product $(\:\cdot\:,\:\cdot\:)_{L^2(M)}$ on 
$\ch^+(M)\oplus\ch^-(M)$. Since $\Vert J\Vert=1$ it
readily follows from this definition that $\Vert JA^\divideontimes JA\Vert=
\Vert A^\divideontimes A\Vert$. The adjoint $\dagger$ and the norm 
$\Vert\:\cdot\:\Vert$ are actually the $*$-operation and norm on the 
particular $C^*$-algebra of bounded linear operators on the particular 
Hilbert space $\ch^+(M)\oplus\ch^-(M)$. Therefore taking into account again 
the uniqueness of $C^*$-algebra norm we also have equalities 
$\Vert A\Vert^2=\Vert A^\dagger A\Vert =r(A^\dagger A)$. Additionally the 
spectral radius always satisfies 
$r(B)=\lim\limits_{k\rightarrow +\infty}\Vert B^k\Vert^{\frac{1}{k}}
\leqq\Vert B\Vert$ which is Gelfand's formula (cf. e.g. 
\cite[Sect. XI.149]{rie-szo}).

After these preparations we can embark upon the proof. On the one hand 
\[r(A^\divideontimes A)=r(JA^\dagger JA)\leqq\Vert JA^\dagger JA\Vert
\leqq\Vert A\Vert^2\:\:\:.\]
On the other hand, for any $\varepsilon >0$ one can find a positive 
integer $k$ such that 
\[\Vert A\Vert^2-\varepsilon =r(A^\dagger A)-\varepsilon 
=r(JA^\divideontimes JA)-\varepsilon\leqq
\Vert (JA^\divideontimes JA)^k\Vert^{\frac{1}{k}}=
\Vert (A^\divideontimes A)^k\Vert^{\frac{1}{k}}\leqq r(A^\divideontimes A)
+\varepsilon\]
therefore, since $\varepsilon >0$ was arbitrary,
\[\Vert A\Vert^2\leqq r(A^\divideontimes A)\:\:\:.\]
We conclude that $r(A^\divideontimes A)=\Vert A\Vert^2$ demonstrating that the 
spectral radius indeed provides us with a norm on $\Omega^2_c(M;\C )$. 
Consequently putting 
\begin{equation}
[[A]]:=\sqrt{r(A^\divideontimes A)}
\label{norma}
\end{equation}
we can complete $V$ with respect to this norm and enrich the $*$-algebra 
$(V,\divideontimes )$ to a $C^*$-algebra $\cb (M)$.

Finally, since diffeomorphisms are unitary i.e., 
$(f^*)^\divideontimes (f^*)={\rm Id}_{\Omega^2_c(M;\C )}$ for all $f\in
{\rm Diff}^+(M)$ we find $[[f^*]]=1$ which means that $f^*\in V\subset\cb (M)$ 
as stated. $\Diamond$ 
\vspace{0.1in}

\begin{remark}\rm From the proof of Lemma \ref{c*lemma} we can also read 
off that although the individual Hilbert space completions 
$\ch^+(M)\oplus\ch^-(M)\supset\Omega^2_c(M;\C )$ might be unitary 
inequivalent, the induced operator norms on the common intersection of the 
individual algebras of bounded linear operators are not only equivalent as 
norms but even {\it numerically equal}. They are commonly given by 
(\ref{norma}). 
\end{remark}

\noindent For a relatively compact open subset 
$\emptyset\subseteqq U\subseteqq M$ a unital $C^*$-algebra 
$\cb (\emptyset )\subseteqq\cb (U)\subseteqq\cb (M)$ 
is defined as the norm-completion of the $\divideontimes$-closed space 
\[\left\{ B\in{\rm End}(\Omega^2_c(M;\C ))\:\left\vert\:[[B]]<+\infty\:,\:
\left[ B\vert_{\Omega^2_c(M\setminus U ;\C )}\right.\:,\:{\rm Diff}^+_U(M)
\right]=0\right\}\]
i.e., $\cb (U)$ consists of operators which commute on 
the subspace $\Omega^2_c(M\setminus U;\C )\subseteqq\Omega^2_c(M;\C )$ 
with the subgroup ${\rm Diff}^+_U(M)\subseteqq{\rm Diff}^+(M)$ consisting 
of all $U$-preserving diffeomorphisms. Since an operator commuting with 
all diffeomorphisms is proportional to the identity, $\cb (\emptyset )
\cong\C\cdot 1$.

Consider the assignment $\{U\mapsto\cb (U)\}_{U\subseteqq M}$ for all 
relatively compact open subsets. Taking into account that if $A\in\cb (U)$ 
then $A\vert_{\Omega^2_c(M\setminus U;\C )}= 
\C\:{\rm Id}_{\Omega^2_c(M\setminus U;\C )}$ and 
$\Omega^2_c(M\setminus V;\C )\subseteqq\Omega^2_c(M\setminus U;\C )$ 
if $U\subseteqq V$ the embedding induces a unit-preserving injective 
homomorphism $e^U_V:\cb (U)\rightarrow\cb (V)$ of local $C^*$-algebras. 
This permits to define $\cb (U)$ for any open 
$\emptyset\subseteqq U\subseteqq M$ and $\cb (M)$ 
as the $C^*$-algebra direct (inductive) limit of these local 
algebras. Henceforth this assignment in fact defines a covariant functor 
from the category of open subsets of $M$ with inclusion into the category of 
unital $C^*$-alegbras with $*$-homomorphisms. However observe that if we 
consider the dual process namely the restriction then elements of these 
local algebras do not behave well because they lack the presheaf property 
in general.

As a consequence of the geometric origin of the global $C^*$-algebra 
$\cb (M)$, it has an interesting sub-$C^*$-algebra ${\mathfrak C}(M)$ if 
$M$ is compact. Indeed, consider the sheaf $\cc_M$ over $M$ whose spaces of 
local sections $\cc (U)$ over open subsets are algebras of local smooth 
bundle (i.e., fiberwise) morphisms
\[C^\infty (U\:;\:{\rm End}(\wedge^2U\otimes_\R\C ))
\:\:\:\:\:\mbox{for all open $U\subseteqq M$}\:\:\:.\]
In contrast to general elements of $\cb (U)$, local sections in $\cc (U)$ 
behave well under restriction due to their presheaf property; i.e., given 
two open subsets $U\subseteqq V$ the restriction map induces a unit-preserving 
injective homomorphism $r^V_U:\cc (V)\rightarrow\cc (U)$ of algebras. 
Although $\cb (U)$ and $\cc (U)$ are not related in general if 
$M$ happens to be compact the space $\cc (M)\subset{\rm 
End}(\Omega^2_c(M;\C ))$ of global sections can be completed with respect 
to (\ref{norma}) to a unital $C^*$-algebra ${\mathfrak C}(M)$ and 
in this case there is an obvious embedding of unital $C^*$-algebras 
${\mathfrak C}(M)\subsetneqq\cb (M)$.
\vspace{0.1in}


\noindent{\it Examples.} The time has come to take a closer look of 
the various operator algebras $\cb (M)$ and $\cc (M)$ (or ${\mathfrak C}(M)$ 
if $M$ is compact) associated to an oriented smooth $4$-manifold $M$ emerging 
through unitary representations of its diffeomorphism group. We will see 
that especially in $4$ dimensions these algebras admit rich physical 
interpretations as follows.

1. Let $(M,g)$ be a $4$-dimensional Riemannian Einstein manifold i.e., 
assume that $g$ is a Riemannian metric on $M$ with Ricci tensor $r_g$ 
satisfying the vacuum Einstein equation $r_g=\Lambda_Mg$ with a cosmological 
constant $\Lambda_M\in\R$. In this special situation the vast symmetry group of 
the original theory reduces to the stabilizer subgroup 
${\rm Iso}^+(M,g)\subsetneqq {\rm Diff}^+(M)$ leaving the geometry $(M,g)$ 
unaffected. In this realm the Riemannian metric together with the orientation 
gives a Hodge operator $*_g:\wedge^2M\rightarrow\wedge^2M$ with 
$*^2_g={\rm Id}_{\wedge^2M}$. This induces a usual real splitting 
\begin{equation}
\wedge^2M=\wedge^+M\oplus\wedge^-M\:\:\:. 
\label{hasitas}
\end{equation}
It is well-known \cite{sin-tho} but from our viewpoint is an 
interesting coincidence that in exactly $4$ dimensions the full Riemannian 
curvature tensor can be regarded as a real linear bundle map 
$R_g:\wedge^2M\rightarrow\wedge^2M$ which as a bundle map decomposes 
i.e., over every point $x\in M$ decomposes like 
\[R_g=\left(\begin{matrix}W^+_g+\frac{s_g}{12} & B_g \\
              B^*_g       & W^-_g+\frac{s_g}{12}
       \end{matrix}\right)\]
with respect to the splitting (\ref{hasitas}). Here the traceless 
symmetric maps $W^\pm_g:\wedge^\pm M\rightarrow\wedge^\pm M$ are the 
(anti)self-dual parts of the Weyl tensor, the diagonal 
$s_g:\wedge^2M\rightarrow\wedge^2M$ is the scalar curvature while 
$B_g:\wedge^+M\rightarrow\wedge^-M$ is the traceless Ricci tensor together 
with its metric adjoint $B^*_g: \wedge^-M\rightarrow\wedge^+M$. 
Observe that the Einstein equation $r_g-\frac{1}{2}s_gg=8\pi T-\Lambda_Mg$ 
exactly says that 
\[\left\{\begin{array}{ll}
B_g&=8\pi T_0\\
s_g&=4\Lambda_M-8\pi\:{\rm tr}_gT
\end{array}\right.\] 
where $T_0$ is the traceless part of the energy-momentum tensor. 
The vacuum $T=0$ is equivalently characterized by the single condition 
$B_g=0$. Indeed, in this case always $T_0=0$ hence if $T\not=0$ then 
matter is present only through its tracial part $(\frac{1}{4}{\rm tr}_gT)g$ 
moreover ${\rm tr}_gT$ is constant by the differential Bianchi identity. 
However by convention such a thing is not called as ``matter'' but rather is 
incorporated into the cosmological constant $\Lambda_M$. Consequently 
looking at the vacuum as being equivalent to the condition $B_g=0$, in the 
case of vacuum $R_g\in C^\infty (M; {\rm End}(\wedge^2M))$ obeys 
(\ref{hasitas}). The pointwise splitting above in addition yields the 
canonical decomposition 
\[\Omega^2_c(M;\C )=\Omega^+_c(M;\C )\oplus\Omega^-_c(M;\C )\]
of the space of $2$-forms into (anti)self-dual forms which is the  
same as decomposing this space into mutually orthogonal maximal definite 
subspaces with respect to the scalar product (\ref{integralas}). Therefore 
in the presence of a metric---which is a way to break the original symmetry 
group ${\rm Diff}^+(M)$ down to a smaller one---there is a 
splitting $\ch^+(M)\oplus\ch^-(M)$ preferred by the curvature $R_g$. 
Switching to our notation we conclude that $R_g\in\cc (M)$ 
satisfies $R_g(\ch^\pm (M))\subseteqq\ch^\pm (M)$. Moreover by the usual 
symmetries of the curvature tensor $R_g$ is self-adjoint for 
(\ref{skalarszorzas}). For clarity we note that this action of for 
example $R_g\in\cc (M)$ on $\ch^+(M)\oplus\ch^-(M)$ is not a Hilbert space 
representation of the $*$-algebra $\cc (M)$ but rather a representation on 
the indefinite space 
$(\ch^+(M)\oplus\ch^-(M),\langle\:\cdot\:,\:\cdot\:\rangle_{L^2(M)})$. 

Therefore we come up with a natural embedding of classical real Riemannian 
(or Lorentzian with complexified curvature) vacuum general relativity 
into a quantum framework:
\vspace{0.1in}

\noindent{\bf C.} {\it The real Riemannian curvature tensor of an 
orientable Riemannian Einstein $4$-manifold $(M,g)$ is a global section 
$R_g\in\cc (M)$ of the sheaf $\cc_M$. The curvature $R_g$ also can be regarded 
as a linear real self-adjoint operator with respect to the
scalar product (\ref{skalarszorzas}) acting on the canonically split Hilbert 
space $\ch^+(M)\oplus\ch^-(M)$ induced by the metric such that $R_g$ obeys this 
splitting. The existence of a metric breaks the 
original symmetry group ${\rm Diff}^+(M)$ down to the finite dimensional group
${\rm Iso}^+(M,g)$ which acts on $\ch^+(M)\oplus\ch^-(M)$ also obeying the 
splitting.} 
\vspace{0.1in}

\begin{remark}\rm Before proceeding further we call attention that---taking 
into account that under mild technical assumptions both the vacuum 
\cite{gas,hal,kow} and the non-vacuum \cite{det} Einstein equations admit 
at least local solutions with prescribed regularity---this classical 
picture is expected to continue to hold at least locally in the following 
sense if one considers more general algebraic curvature tensors. Given a 
connected oriented smooth $4$-manifold $M$ with a point $x\in M$ it is 
known that if a global algebraic curvature tensor $R_M\in\cc (M)$ 
satisfies some technical conditions in $x$ (formulated for example in 
\cite{gas, hal, kow}), then there exists at least a local Riemannian 
Einstein metric $g_U$ on an open subset $x\in U\subseteqq M$ with the 
property $R_{g_U}\vert_x=R_M\vert_x=R_x$ i.e., the two curvature tensors 
coincide at least in $x$. Apparently we can pick a countable 
collection of distinguished points of this kind such that the corresponding 
open subsets comprise an open covering of $M$ hence endowing $M$ with a 
``patchwork structure'' of local Einstein metrics. 
\end{remark}

2. Next we take a departure from classical general relativity 
and explore the quantum regime. Of course the trouble is how to 
describe a generic bounded linear operator $Q\in\cb (M)$ in terms of a 
geometric linear operator $R\in\cc (M)\cap\cb (M)$. Our quantum 
instinct tells us that a truely quantum operator should be constructed 
by somehow smearing geometric operators over regions in $M$. This 
instinct will be justified by the famous Schwartz kernel theorem applied below.

Fix a geometric operator $R\in\cc (M)\cap\cb (M)$ and a point $x\in M$. Then 
on any $2$-form $\omega\in\Omega^2_c(M;\C )$ its action can be expressed in a 
fully local form $(R\omega )_x=R_x\omega_x$. We can generalize this 
as follows. Pick finitely many distinct further points $y_1,\dots, 
y_{n(x)}\in M$ where $n(x)\in\N$ may depend on $x\in M$. Consider 
diffeomorphisms $f_{y_i}\in{\rm Diff}^+(M)$ such that $f_{y_0}={\rm Id}_U$ 
hence $f_{y_0}(x)=x$ moreover $f_{y_i}(x)=y_i$ for $i=1,\dots, n(x)$. An 
operator $Q\in\cb (M)$ out of $R\in\cc (M)\cap\cb (M)$ and 
$f_{y_0},\dots,f_{y_{n(x)}}\in{\rm Diff}^+(M)$ is constructed such that on 
vectors $\omega\in\Omega^2_c(M;\C )$ forming a dense subset has the shape 
\begin{equation} 
(Q\omega )_x:=\sum\limits_{i=0}^{n(x)}f^*_{y_i}(R\omega ) 
=R_x\omega_x+\sum\limits_{i=1}^{n(x)}f^*_{y_i}(R\omega )\:\:\:. 
\label{approximalas1} \end{equation} Note that this linear operator is 
not local in the sense that its effect on $\omega_x$ depends 
not only on $R_x$ and $\omega_x$ but on the value of $R$ and $\omega$ in 
further distant points $y_1,\dots,y_{n(x)}\in M$ as well. The question 
arises how to generalize this construction for countable or even 
uncountable infinite sums. For all points $y\in M$ pick up unique 
diffeomorphisms $f_y\in{\rm Diff}^+(M)$ such that $f_y(x)=y$ and 
$f_x={\rm Id}_M$. Then for all $\omega\in\Omega^2_c(M;\C )$ the assignment 
$y\mapsto f^*_y(R\omega )$ gives a function from $M$ into 
$\wedge^2_xM\otimes_\R\C$. Suppose we can integrate it against a complex 
{\it measure} $\mu_x$ on $M$ what we write as $\int_{y\in M}f^*_y(R\omega )
\dd\mu_x(y)$.  Such a measure can be constructed from a {\it 
double $2$-form} $K$ i.e., a section of the bundle $(\wedge^2M\otimes_\R\C )
\times(\wedge^2M\otimes_\R\C )$ over $M\times M$ regarding it as a 
``kernel function''. In other words for all $x\in M$ and a $2$-form $\omega$ 
we put 
\[\int\limits_{y\in M}f^*_y(R\omega )\dd\mu_x(y):= 
\int\limits_{y\in M}K_{x,y}\wedge 
(R\omega )_y\in\wedge^2_xM\otimes_\R\C\:\:\:.\] 
Consequently the appropriate way to generalize the discrete formula 
(\ref{approximalas1}) is to set 
\[(Q\omega )_x:=\int\limits_{y\in M}K_{x,y}\wedge (R\omega )_y\:\:\:.\] 
Of course in order this integral to make sense we have to specialize the 
precise class of these ``kernel functions''. We shall not do it here but note 
that the more singular the kernel is, the more general is the resulting bounded 
linear operator. The general situation is controlled by the Schwartz 
kernel theorem: non-tempered distributional double $2$-forms $K\in\cd '
(M\times M\:;\:(\wedge^2M\otimes_\R\C )\times (\wedge^2M\otimes_\R\C ))$ 
give rise to bounded linear operators $Q$ via 
$\langle\alpha ,Q\beta\rangle_{L^2(M)}=
(K, \alpha\otimes (R\beta) )_{M\times M}$ where this latter bracket is the 
pairing between dual spaces (cf. e.g. \cite[Vol. I Sect. 4.6]{tay}) 
and all bounded linear operators arise this way with suitable kernels.

\vspace{0.1in}

\noindent{\bf Q.} {\it Over a connected oriented smooth $4$-manifold $M$ a 
generic element $Q\in\cb (M)$ always can be constructed from a geometric one 
$R\in\cc (M)\cap\cb (M)$ by a smearing procedure provided by the Schwartz 
kernel theorem. In this general situation no pointwisely given geometric 
object has a meaning because the original symmetry group ${\rm Diff}^+(M)$ is 
unbroken. This is in accord with the physical expectations.}
\vspace{0.1in}

\noindent We have completed the exploration of the elements of $\cc (M)$ 
and $\cb (M)$.


\section{Gravity as an algebraic quantum field theory}
\label{three}


Before proceeding further let us summarize the situation we have reached 
in Sect. \ref{two}. To a smooth oriented $4$-manifold $M$ one can attach a 
sheaf $\cc_M$ whose global sections $\cc (M)$ contains algebraic curvature 
tensors. $\cc (M)$ often can be completed to a $C^*$-algebra ${\mathfrak 
C}(M)$. Classical solutions of the vacuum Einstein equations i.e., 
classical real Riemannian (or Lorentzian with complexified curvature) 
Einstein manifolds $(M,g)$ can be characterized by the fact that their 
curvature operators obey the canonical splitting $\Omega^+_c(M;\C ) 
\oplus\Omega^-_c(M;\C )\subset\ch^+(M)\oplus\ch^-(M)$ and this completion 
equipped with an indefinite scalar product carries a representation of 
$\cc (M)$ or even ${\mathfrak C}(M)$ and a unitary one of ${\rm 
Diff}^+(M)$. Therefore one is tempted to look at curvature operators as 
local quantum observables in a quantum field theory possessing a huge 
symmetry group coming from diffeomorphisms. We make these observations 
more formal by constructing something which resembles an {\it algebraic 
quantum field theory} in the sense of \cite{haa}. For this aim we need a 
``net'' or a ``co-presheaf'' of local algebras on $M$ i.e., a functorial 
assignment $O\mapsto\fu (O)$ attaching $C^*$-algebras $\fu (O)$ to open 
subsets $\emptyset\subseteqq O\subseteqq M$ such that the basic axioms of 
this theory having still meaning in our more general context should be 
satisfied.

Recall that the space of local smooth complexified $(0,4)$-type algebraic 
curvature tensors over $M$ is $C^\infty (M;(S^2\wedge^2M\cap{\rm 
Ker}\:b)\otimes_\R\C )$ where $b:C^\infty (M; (\wedge^1M)^{\otimes 4})
\rightarrow C^\infty (M; (\wedge^1M)^{\otimes 4})$ is the usual 
algebraic Bianchi map. Making use of a metric i.e., pseudo-Euclidean 
structures on the fibers, the corresponding $(2,2)$-type algebraic 
curvature tensors fulfill a subspace of $C^\infty (M; {\rm End}(\wedge^2 
M\otimes_\R\C ))$. However now we lack any preferred metric hence only the 
whole endomorphism space is at our disposal. Consider therefore 
${\rm End}(\Omega^2_c(M;\C ))$, the adjoint operation $\divideontimes$ with 
respect to (\ref{integralas}) and the norm (\ref{norma}) given by the 
spectral radius. Take compactly supported complex bundle morphisms 
$R\in C_c^\infty (M;{\rm End}(\wedge^2M \otimes_\R\C ))$ and {\it real} 
vector fields $X\in C_c^\infty (M;TM)$ with the associated Lie derivative 
$L_X$. Then ${\rm e}^R$ as well as ${\rm e}^{L_X}$ have finite norm 
(\ref{norma}). Fix a relatively compact open subset 
$\emptyset\subseteqq U\subseteqq M$ and let $\fu (U)$ be the unital 
$C^*$-algebra generated by the operators ${\rm e}^R, {\rm e}^{L_X}$ which 
commute on $\Omega^2_c(M\setminus U;\C )\subset\Omega^2_c(M;\C )$ 
with the subgroup ${\rm Diff}^+_U(M)\subset {\rm Diff}^+(M)$ consisting of 
$U$-preserving diffeomorphisms. I.e., $\fu (U)$ arises as the 
norm-closure for (\ref{norma}) of the $\divideontimes$-closed subspace 
\[\left\langle {\rm e}^R, {\rm e}^{L_X}\:
\:\left\vert\:\left[ {\rm e}^R\vert_{\Omega^2_c(M\setminus U;\C )}\:,
\:{\rm Diff}^+_U(M)\right] =0\:,\: 
\left[{\rm e}^{L_X}\vert_{\Omega^2_c(M\setminus U;\C )}
\:,\:{\rm Diff}^+_U(M)\right] =0\right.\right\rangle\:\:\:.\] 
By construction $\C\cdot 1\cong\fu (\emptyset )
\subseteqq\fu (U)\subseteqq\fu (V)$ if $\emptyset\subseteqq U\subseteqq V$ 
therefore, as usual, the global algebra $\fu (M)$ is constructed (if $M$ is 
non-compact) as the $C^*$-algebra direct (inductive) limit of these local 
algebras.

\begin{definition} The algebra $\fu (U)$ is called the {\rm local 
generalized CCR algebra} of local quantum observables while $\fu (M)$ is 
the {\rm global generalized CCR algebra} of $M$. 
\label{ccr}
\end{definition} 

\begin{remark}\rm 1. This definition of local quantum 
observables stems from the physical intuition that on remote 
localized states local operations should commute with localization-preseving 
symmetries. 

2. $\fu (U)$ contains a usual CCR algebra at least when $U\subseteqq M$ is 
a coordinate ball. Pick self-adjoint local endomorphisms $R$ and local 
vector fields $X$ with $L_X$ being self-adjoint. Since $X$ is real then 
${\rm e}^{L_X}$ is a diffeomorphism which is unitary hence $L_X$ is 
self-adjoint. Consider the maximal subspace of those self-adjoint elements 
which either commute: $[R_1, R_2]=0$, $[L_{X_1}, L_{X_2}]=0$, $[R,L_X]=0$ 
or are canonically conjugate to each other i.e., $[R,L_X]=c\cdot 1$ with 
$c\in\C$. Then the sub-$C^*$-algebra in $\fu (U)$ generated by the 
corresponding unitary operators ${\rm e}^R, {\rm e}^{L_X}$ form a usual 
CCR algebra; $R$ and $L_X$ play the role of the position operator ${\bf Q}$ 
and its canonically conjugate momentum operator ${\bf P}$, 
respectively. This standard CCR algebra within $\fu (U)$ describes the 
``free graviton part'' while the rest of $\fu (U)$ the ``self-interacting 
part'' of this theory. This justifies in some extent why we expect to 
construct something like a ``quantum theory''. 
\end{remark}

\noindent Putting things together then let us consider the algebraic 
quantum field theory defined by the assignment
\[U\longmapsto\fu (U),\:\:\:\:\:\mbox{$U\subseteqq M$ is 
relatively compact open.}\]
Moreover $\fu (M)$ is taken to be the $C^*$-algebra direct (inductive) 
limit of the $\fu (U)$'s as usual. Note that the formulation of this theory 
rests only on the smooth structure on $M$ hence does not refer to any metric 
on $M$ for instance. A Hilbert space $\ch^+(M)\oplus\ch^-(M)$ carries an 
action of all $\fu (U)$'s from the left and a unitary representation with 
respect to $\langle\:\cdot\:,\:\cdot\:\rangle_{L^2(M)}$ of 
${\rm Diff}^+(M)$ from the right. Elements of the algebra $\fu (U)$ are the 
{\it local quantum observables} and those of the group ${\rm Diff}^+(M)$ are
the {\it symmetry transformations}. The {\it states} are continuous 
normalized positive linear functionals on $\fu (M)$ and the 
{\it expectation value} of $B\in\fu (M)$ in the state $\Phi$ is 
$\Phi (B)\in\C$. 

Now we introduce the concept of a ``quantum gravitational field'' in the 
standard way. 

\begin{definition} Let $M$ be a connected oriented smooth $4$-manifold. 
Take a local generalized CCR algebra $\fu (U)$ generated by ${\rm e}^R$'s 
and ${\rm e}^{L_X}$'s as above. For a differentiable $1$-parameter subgroup 
$\{A_t\}_{t\in\R}\subset\fu (U)$ with $A_0=1\in\fu (U)$ a local 
observable of the infinitesimal form
\[Q:=\left.\frac{\dd A_t}{\dd t}\right\vert_{t=0}\in T_1\fu (U)\]
is a called a {\rm local quantum gravitational field} on $U\subseteqq M$. 

Take any split Hilbert space $\ch^+(M)\oplus\ch^-(M)$ 
containing maximal definite orthogonal subspaces (note that this breaks 
the diffeomorphism symmetry). The off-block\-diagonal part of $Q$ is the 
{\rm material content of the local quantum gravitational field relative to 
the splitting}. In particular $Q$ is called a {\rm local quantum vacuum
gravitational field relative to the splitting} if its material content
relative to the splitting vanishes i.e., 
$Q(\ch^\pm (M)\cap D)\subseteqq\ch^\pm (M)$ at least on a dense subset 
$D\subseteqq \ch^+(M)\oplus\ch^-(M)$.
\label{kvantum}
\end{definition}

\noindent Now we turn to the representation theory of the global algebra 
$\fu (M)$. As usual this global CCR algebra of observables admits an 
abundance of non-equivalent representations therefore an important task is 
to single out those which possess some---either mathematical or 
physical---significance. 

Firstly we construct what will be referred to as the 
{\it tautological representation} having probably a mathematical 
relevance only.

\begin{theorem} $M$ itself gives rise to a faithful 
and irreducible so-called {\rm tautological representation} $\pi_M$ of 
$\fu (M)$ on a Hilbert space $\ch_M$. It also carries a unitary 
representation $U_M$ of the group ${\rm Diff}^+(M)$. A vector $v\in\ch_M$ 
satisfies $U_M(v)=v$ if and only if $v=0$ (``no vacuum'').

As a consequence to $M$ always a von Neumann algebra 
${\mathfrak R}(M):=(\pi_M(\fu (M)))''$ can be attached canonically.
\label{diff-top-reprezentacio}
\end{theorem}
 
\noindent{\it Proof.} Referring back to Lemma \ref{c*lemma} we improve 
$\fu (M)$ itself to a Hilbert space $\ch_M$ on which $\fu (M)$ acts from 
the left. Recall that $\fu (M)$ has a norm given by the spectral radius 
(\ref{norma}). We want to demonstrate that this norm $[[\:\cdot\:]]$ actually 
comes from a positive definite non-degenerate Hermite scalar product 
$(\:\cdot\:,\:\cdot\:)_M$. This will also yield that the Hilbert space 
completion $\ch_M$ of $\fu (M)$ will actually coincide with $\fu (M)$ 
i.e., $\ch_M$ will arise simply by putting this scalar product onto $\fu (M)$. 

Define a map from $\fu (M)^\R\times\fu (M)^\R$ into $\R$ by 
differentiating $T\mapsto [[T]]^2$ at the unit $1\in\fu (M)$ as follows: 
\[\fu (M)^\R\times\fu (M)^\R\ni (A,B)\longmapsto\frac{1}{4}
({\rm D}[[\:\cdot\:]]^2)_1(A^\divideontimes B+B^\divideontimes A)\in\R\:\:\:.\]
Properties of the norm ensure us that this derivative exists and the map 
is symmetric and $\R$-bilinear. Take any particular 
Hilbert space $\ch^+(M)\oplus\ch^-(M)$ from the proof of Lemma \ref{c*lemma}. 
Recall the equality $[[A]]=\Vert A\Vert$ for all $A\in\fu (M)$ where 
$\Vert\:\cdot\:\Vert$ is the usual operator norm on this Hilbert space 
satisfying (\ref{opinorm}). Then 
\[\frac{1}{2}\:{\rm D}\left(\frac{\Vert (\:\cdot\:)v\Vert^2_{L^2(M)}}
{\Vert v\Vert^2_{L^2(M)}}\right)_{\!\!1}(A^\divideontimes A)
=\frac{ {\rm Re}(A^\divideontimes Av\:,\:v)_{L^2(M)}}{\Vert v\Vert^2_{L^2(M)}}=
\frac{{\rm Re}(JAv\:,\:AJv)_{L^2(M)}}{\Vert v\Vert^2_{L^2(M)}}\]
hence these derivatives also exist and taking their supremum with respect to 
$v\in\ch^+(M)\oplus\ch^-(M)$ gives $\Vert A\Vert^2$. Consequently  
\[\frac{1}{2}({\rm D}[[\:\cdot\:]]^2)_1(A^\divideontimes A)=
\frac{1}{2}{\rm D}\left(\sup\limits_{v\not=0}\frac{\Vert
(\:\cdot\:)v\Vert^2_{L^2(M)}}
{\Vert v\Vert^2_{L^2(M)}}\right)_{\!\!1}\!\!(A^\divideontimes A)
=\sup\limits_{v\not=0}\:\frac{1}{2}{\rm D}\left(\frac{\Vert
(\:\cdot\:)v\Vert^2_{L^2(M)}}
{\Vert v\Vert^2_{L^2(M)}}\right)_{\!\!1}\!\!(A^\divideontimes A) 
=\Vert A\Vert^2\:\:\:.\]
This shows that $\frac{1}{2}({\rm D}[[\:\cdot\:]]^2)_1(A^\divideontimes A)
=\Vert A\Vert^2\geqq 0$ and equality holds if and only if $A=0$. 
Therefore $(A,B)\mapsto\frac{1}{4}({\rm D}[[\:\cdot\:]]^2)_1
(A^\divideontimes B+B^\divideontimes A)$ is a real non-degenerate scalar 
product on $\fu (M)^\R$ with induced 
norm $[[\:\cdot\:]]$. The norm satisfies $[[A]]=[[\ii A]]$ over $\fu (M)$ as 
well therefore putting
\[(A\:,\:B)_M:=\frac{1}{2}\left( [[A+B]]^2-[[A]]^2-[[B]]^2\right) +
\frac{\ii}{2}\left( [[\ii A+B]]^2-[[\ii A]]^2-[[B]]^2\right)\] 
gives rise to a non-degenerate Hermitian scalar product on $\fu (M)$. 
In other words $\fu (M)$ as a complete normed space has the 
further structure of a Hilbert space $\ch_M$ and $\fu (M)$ acts on it(self) 
from the left yielding a faithful irreducible representation $\pi_M$ 
i.e., $\pi_M(A)B:=AB$ for all $A\in\fu (M)$, $B\in\ch_M=\fu (M)$.

Since by construction ${\rm Diff}^+(M)\subset\fu (M)$ we also 
obtain a unitary representation $U_M(f):=\pi_M(f^*)$ and 
via part (ii) of Lemma \ref{vakuum} obviously $v=0$ is the 
only invariant vector under $U_M$ as stated. $\Diamond$
\vspace{0.1in}

\noindent Secondly, in a quantum field theory the algebra of quantum 
observables must possess positive mass and energy representations. Let us 
therefore construct some representations $\pi_{\Sigma ,\omega}$ of our 
global algebra $\fu (M)$ what we will call {\it positive mass 
representations}. When doing this we touch upon the problem of 
gravitational mass and energy which is probably the most painful part of 
current general relativity \cite{sza}.

\begin{theorem}
Take an oriented closed surface $\Sigma$. Let 
$(\Sigma ,p_1,\dots ,p_n )$ denote a generic smooth immersion 
$i:\Sigma\looparrowright M$ where the points $p_1,\dots,p_n\in\Sigma$ are the 
preimages of the double points of this immersion. Moreover take 
any closed $\omega\in\Omega_c^2(M;\C )$. Assume that 

\begin{itemize}  

\item[(i)] $\frac{1}{2\pi\ii}\int_\Sigma \omega =1$;

\item[(ii)] $\omega$ is non-degenerate along $\Sigma$ and for all complex 
structures $C=C(\Sigma )$ on $\Sigma$ there exist positive 
definite unitary holomorphic vector bundle 
structures on the vector bundle $E:=TM\otimes_\R\C\vert_C$ over $C\subset M$ 
compatible with $\omega$ such that $\dim_\C H^0(C;\co (E))=4$.
\end{itemize}

\noindent Then $(\Sigma ,p_1,\dots,p_n,\omega )$ gives rise to a 
so-called {\rm positive mass representation} $\pi_{\Sigma ,\omega}$ of 
$\fu (M)$ on a Hilbert space $\ch_{\Sigma ,\omega}$ as follows:

\begin{itemize}

\item[(i)] $\ch_{\Sigma,\omega}$ also carries a unitary representation 
$U_{\Sigma ,\omega}$ of the group ${\rm Diff}^+(M)$. A vector 
$v\in\ch_{\Sigma ,\omega}$ satisfies $U_{\Sigma ,\omega}(f)v=v$ for all 
$f\in{\rm Diff}^+(M)$ if and only if $v=0$ (``no vacuum'');

\item[(ii)] On a dense subset of states $0\not= [A]\in\ch_{\Sigma 
,\omega}$ a complex $4$-vector $P_{C,\omega ,A}\in H^0(C;\co (E))$ can be 
defined together with its length $m_{C,\omega ,A}:=
\Vert P_{C,\omega , A}\Vert_{L^2(C)}\geqq 0$ with respect to a natural 
Hermitian scalar product $(\:\cdot\:,\:\cdot\: )_{L^2(C)}$ on 
$C^\infty (C;E)$. It has the porperty that if 
$[1]\in\ch_{\Sigma ,\omega}$ is a state corresponding to vanishing 
algebraic curvature $R=0$ then $P_{C,\omega ,1}=0$ hence 
$m_{C,\omega ,1}=0$.
\end{itemize}

\label{pozitiv-tomegu-reprezentacio}
\end{theorem}

\noindent {\it Proof.} (i) A continuous functional $\Phi_{\Sigma ,\omega}: 
\fu (M)\rightarrow\C$ is defined by extending continuously the map   
\[A\longmapsto\Phi_{\Sigma,\omega}(A):=
\frac{1}{2\pi\ii}\int\limits_\Sigma A\omega\in\C\]
from ${\rm End}(\Omega^2_c(M;\C ))\cap\fu (M)$. By assumption (i) 
$\Phi_{\Sigma ,\omega}(1^\divideontimes 1)=1$ hence $\Phi_{\Sigma 
,\omega}(A^\divideontimes A)>0$ if $\Vert A-1\Vert <\varepsilon$. Since any 
$B\in\fu (M)$ can be written as $B=cA$ with $c\in\C$ we see that 
$\Phi_{\Sigma ,\omega}(B^\divideontimes B)=\vert c\vert^2\Phi_{\Sigma ,\omega}
(A^\divideontimes A)\geqq 0$ consequently $\Phi_{\Sigma ,\omega}$ is 
continuous, normalized and positive. Therefore the GNS construction 
applies and yields a corresponding representation of the $C^*$-algebra 
$\fu (M)$. Recall that this goes as follows. One has the induced 
left-multiplicative Gelfand ideal 
$I_{\Sigma ,\omega}:=\{A\in\fu (M)\:\vert\: \Phi_{\Sigma 
,\omega}(A^\divideontimes A)=0\}\subset\fu (M)$. The functional provides us 
with a well-defined positive definite scalar product 
$([A], [B])_{\Sigma ,\omega}:=\Phi_{\Sigma, \omega}(A^\divideontimes B)$ on 
$\fu (M)/I_{\Sigma ,\omega}$ with $A\in [A], B\in [B]$ where 
$[A]:=A+I_{\Sigma ,\omega}$, etc. Making use of this scalar product one 
completes $\fu (M)/I_{\Sigma ,\omega}$ to a Hilbert space 
$\ch_{\Sigma ,\omega}$ and then lets $\fu (M)$ act from the left by the 
continuous extension of $\pi_{\Sigma ,\omega}(A)[B]:=[AB]$ from 
$\fu (M)/I_{\Sigma ,\omega}$ to $\ch_{\Sigma ,\omega}$. Since the whole 
construction is acted upon equivariantly by ${\rm Diff}^+(M)$ 
(i.e., all the $M$, $\fu (M)$ and $\ch_{\Sigma ,\omega}$ carry 
induced actions of the diffeomorphism group) two representations 
$\pi_{\Sigma ,\omega}$ and $\pi_{f(\Sigma ),f^*\omega}$ are considered 
to be {\it identical} and the whole set of these representations will be 
denoted by 
$\pi_{\Sigma ,\omega}$ with Hilbert space $\ch_{\Sigma ,\omega}$. In fact two 
representations $\pi_{\Sigma_1,\omega_1}$ and $\pi_{\Sigma_2,\omega_2}$ are 
{\it unitary equivalent} if and only if there is a positive real number 
$a\in\R^+$ such that $\Phi_{\Sigma_2,\omega_2}=a\Phi_{\Sigma_1,\omega_1}$ 
hence in particular if exists an element $f\in{\rm Diff}^+(M)$ satisfying 
$\Sigma_2=f(\Sigma_1)$ and $\omega_2=f^*\omega_1$; consequently our 
identification is consistent from a representation-theoretic viewpoint as well.

In usual Poincar\'e-invariant quantum field theory the Hilbert space 
carries a unitary representation of the space-time symmetry group. 
Here the ``space-time symmetry transformations'' are all 
the diffeomorphisms hence in our algebraic quantum field theory the 
corresponding infinitesimal transformations are the Lie derivatives with 
respect to vector fields. We construct a unitary 
representation $U_{\Sigma ,\omega}$ of ${\rm Diff}^+(M)$ on 
$\ch_{\Sigma ,\omega}$ from the left as follows. An element 
$f\in{\rm Diff}^+(M)$ arises as limits of products of diffeomorphisms whose 
infinitesimal generators are compactly supported real vector 
fields $X\in C^\infty_c(M;TM)$. Recalling the 
construction of $\fu (M)$ we see that if one looks at the pullback $f^*$ 
induced by $f\in {\rm Diff}^+(M)$ as a linear operator on 
$\Omega^2_c(M;\C )$ then $f^*\in\fu (M)$ and it is a unitary element. 
Therefore put $U_{\Sigma ,\omega}(f):=\pi_{\Sigma ,\omega}(f^* )$ for 
$f\in {\rm Diff}^+(M)$. This representation is indeed unitary because 
\[(U_{\Sigma ,\omega}(f)[A], U_{\Sigma ,\omega}(f)[B])_{\Sigma ,\omega}
=([f^*A],[f^*B])_{\Sigma 
,\omega}= \Phi_{\Sigma ,\omega}((f^*A)^\divideontimes (f^* B))
=\Phi_{\Sigma ,\omega}(A^\divideontimes B)=([A],[B])_{\Sigma ,\omega}\:.\]
The representation $U_{\Sigma ,\omega}: {\rm Diff}^+(M)\rightarrow
{\rm U}(\ch_{\Sigma ,\omega})$ has a complexified infinitesimal version 
\[u_{\Sigma ,\omega}: C^\infty_c(M; TM\otimes_\R\C )\cong 
{\rm Lie}({\rm Diff}^+(M))\otimes_\R\C\longrightarrow
{\mathfrak u}(\ch_{\Sigma ,\omega})\otimes_\R\C\cong
{\rm End}(\ch_{\Sigma ,\omega})\]
whose matrix elements on the dense subspace 
$D:=(\fu (M)\cap{\rm End}(\Omega^2_c(M;\C )))/I_{\Sigma ,\omega}
\subset\ch_{\Sigma ,\omega}$ look like 
\[([A], u_{\Sigma ,\omega}(X)[B])_{\Sigma ,\omega}=
\lim\limits_{t\rightarrow 0}\:\Phi_{\Sigma ,\omega}
\left( A^\divideontimes\:\frac{{\rm e}^{tX}B-B}{t}\right) =
\frac{1}{2\pi\ii}\int\limits_\Sigma A^\divideontimes L_XB\omega\:\:\:.\] 
For real vector fields we have a more geometric description: 
if $\{ f_t\}_{t\in\R}\subset {\rm Diff}^+(M)$ is a $1$-parameter subgroup 
for $X\in C^\infty_c(M; TM)$ then 
$u_{\Sigma ,\omega}(X)[A]\in\ch_{\Sigma ,\omega}$ satisfies 
\[\lim\limits_{t\rightarrow 0}\left\Vert u_{\Sigma ,\omega}(X)[A]
-\left[\frac{f^*_tA-A}{t}\right]\right\Vert_{\Sigma ,\omega}=0\] 
consequently the $u_{\Sigma ,\omega}(X)$'s are indeed 
the complexified infinitesimal generators of ${\rm Diff}^+(M)$ in the 
representation $U_{\Sigma ,\omega}$. It follows from part (i) of 
Lemma \ref{vakuum} that the only invariant vector under 
${\rm Diff}^+(M)$ is $0\in\ch_{\Sigma ,\omega}$.

(ii) In usual Poincar\'e-invariant quantum field theory a $4$ dimensional 
commuting set of infinitesimal space-time symmetries 
are regarded as infinitesimal translations; the 
corresponding operators are interpreted as energy-momentum operators acting 
on the Hilbert space of the theory. However in general one cannot find a 
distinguished $4$ dimensional commuting subspace which could be called 
as ``infinitesimal translations''. To overcome this difficulty we will follow 
Dougan and Mason \cite{dou-mas} (or \cite[Chapter 8]{sza}) to find a 
distinguished subspace of vector fields by holomorphy. 

Consider $E:=TM\otimes_\R\C\vert_\Sigma$ satisfying ${\rm rk}_\C E = 4$ and 
${\rm deg}E=0$. The $2$-form $\omega$ can also be used to construct a 
Hermitian metric on it. Indeed, a Hermitian form on $M$ is defined by 
$g(X,Y) :=\frac{1}{2}\left(\omega 
(\overline{X},\ii Y)-\omega (\overline{\ii X}, Y)\right)$ 
for all $X,Y\in C^\infty (M;TM\otimes_\R\C )$. By
assumptions (ii) in the theorem its restriction makes $E$ into a smooth 
positive definite unitary vector bundle $(E,g)$ over $\Sigma$. Take a 
connection $\nabla_E:C^\infty (\Sigma ; E)\rightarrow 
C^\infty (\Sigma ; E\otimes_\C\wedge^1\Sigma )$ satisfying $\nabla_Eg = 0$ 
which means that it is compatible with the unitary structure. Picking any 
complex structure on $\Sigma$ we can identify it with a compact Riemann 
surface $C = C(\Sigma )$. The $(0,1)$-part $\overline{\partial}_E$  
of the connection endows $(E,g)$ with the structure 
of a unitary holomorphic vector bundle over $C$. Its finite dimensional 
subspace of holomorphic sections is denoted by 
$H^0(C;\co (E))$. The Riemann--Roch--Hirzebruch theorem gives
$\dim_\C H^0(C;\co (E))\geqq 4(1-{\rm genus}(C))$ but by assumptions 
(ii) in the theorem this vector space is supposed to be precisely $4$ 
dimensional. It also follows that the Hermitian $L^2$ scalar 
product $(X,Y)_{L^2(C)}:=\frac{1}{2\pi\ii}\int_Cg(X,Y)\omega$ on 
$C^\infty (C;E)$ is positive definite. A simple choice for $E$ can 
be the holomorphically trivial bundle $C\times\C^4$.

We already have seen that the expectation value of any vector field $X$ 
on $M$ is well defined for a dense subset of 
vectors $0\not=v\in D\subset\ch_{\Sigma ,\omega}$ and looks like 
$\frac{(v\:,\: u_{\Sigma ,\omega}(X)v)_{\Sigma ,\omega}}
{\Vert v\Vert^2_{\Sigma ,\omega }}\in\C .$ However
\[([A], u_{\Sigma ,\omega}(X)[A])_{\Sigma ,\omega}=
\frac{1}{2\pi\ii}\int\limits_\Sigma A^\divideontimes L_XA\omega\]
by construction, therefore 
\[\frac{([A]\:,\: u_{\Sigma ,\omega}(X)[A])_{\Sigma ,\omega}}
{\Vert [A]\Vert^2_{\Sigma ,\omega }}=\frac{\int\limits_\Sigma 
A^\divideontimes L_XA\omega}{\int\limits_\Sigma A^\divideontimes A\omega}\]
which is complex linear in $X\in C^\infty_c(M;TM\otimes_\R\C )$. 
Let $N_\varepsilon (C)\subset M$ be a small tubular neighbourhood of 
$C\subset M$; we can suppose that it is a $B^2_\varepsilon$-bundle over 
$C$ hence put $N_0(C):=C$. Take a linear functional on 
$C^\infty (N_\varepsilon (C);TN_\varepsilon (C)\otimes_\R\C)$ by setting 
\[P_{\varepsilon ,C,\omega ,A}(X\vert_{N_\varepsilon (C)})
:=\frac{\int\limits_CA^\divideontimes L_XA\omega}{\int\limits_C 
A^\divideontimes A\omega}\:\:\:\]
and then on $C^\infty (C;E)$ by
\begin{equation}
P_{C,\omega ,A}(X\vert_C):=\lim\limits_{\varepsilon\rightarrow 0}
P_{\varepsilon ,C,\omega ,A}(X\vert_{N_\varepsilon (C)})\:\:\:.
\label{dougan--mason}
\end{equation}
A vector field $X\in C^\infty_c(M;TM\otimes_\R\C )$ is called a 
{\it quasilocal infinitesimal translation along $\Sigma$} if $X\vert_C\in 
H^0(C;\co (E))\subset C^\infty (C;E))$ and (\ref{dougan--mason}) 
gives rise to a well-defined dual vector 
$P_{C,\omega ,A}\in (H^0(C;\co (E)))^*$. This $P_{C,\omega ,A}$ is called the 
{\it quasilocal energy-momentum $4$-vector along $\Sigma$} of the state 
$[A]\in\ch_{\Sigma ,\omega}$. By the aid of the scalar product $(\:\cdot\:,
\:\cdot\:)_{L^2(C)}$ we identify $(H^0(C;\co (E)))^*$ with 
$\overline{H^0(C;\co (E))}$ therefore we can suppose that 
$P_{C,\omega ,A}\in H^0(C;\co (E))$ by putting 
$(\overline{P}_{C,\omega ,A}, X\vert_C)_{L^2(C)}:=P_{C,\omega ,A}(X\vert_C)$. 
By assumptions (ii) of the theorem $P_{C,\omega ,A}$ is indeed a complex 
$4$-vector. Its length looks like 
\[\Vert P_{C,\omega ,A}\Vert^2_{L^2(C)}=
\frac{1}{2\pi\ii}\int\limits_Cg\left( 
P_{C,\omega, A}\:,\:P_{C,\omega, A}\right)\omega =
\frac{1}{2\pi\ii}\int\limits_C 
\overline{P}_{C,\omega ,A}\wedge *_gP_{C,\omega ,A}=
\frac{1}{2\pi\ii}\int\limits_C\vert 
P_{C,\omega ,A}\vert^2_g\omega\]
and we call the number $m_{C,\omega ,A}:=\Vert P_{C,\omega ,A}\Vert_{L^2(C)}
\geqq 0$ the {\it mass} of the the state $[A]\in\ch_{\Sigma ,\omega}$. 

Finally, the ``semiclassical gravitational vacuum'' defined by $R=0$ along 
$M$ is represented by the state $[{\rm e}^0]\in\ch_{\Sigma ,\omega}$. However 
${\rm e}^0=1\in\fu (M)$ hence $[{\rm e}^0]=[1]$. Consequently with some 
$\varepsilon > 0$ for any quasilocal infinitesimal translation $X$ along 
$\Sigma$ we find 
\[P_{\varepsilon, C,\omega ,1}(X\vert_{N_\varepsilon(C)})=
\frac{1}{2\pi\ii}\int\limits_C L_X\omega =
\frac{1}{2\pi\ii}\int\limits_C\left(\iota_X\dd\omega +\dd(\iota_X\omega )
\right) =0\]
because both $C$ and $\omega$ are closed by assumption. Therefore 
taking $\varepsilon\rightarrow 0$ the expression 
(\ref{dougan--mason}) yields $P_{C,\omega ,1}=0$ that is, this state has 
zero quasilocal energy-momentum hence mass as expected. $\Diamond$ 
\vspace{0.1in}

\noindent\begin{remark}\rm 1. The formula (\ref{dougan--mason})
for the quasilocal energy-momentum formally remains meaningful for quantum
gravitational fields introduced in Definition \ref{kvantum}. Hence the
corresponding quantities $P_{C,\omega ,Q}$ and 
$m_{C,\omega ,Q}$ are interpreted as the quasilolcal energy-momentum    
$4$-vector and the mass of a quantum gravitational field $Q$. Among local  
quantum gravitational fields one can recognize classical curvature tensors
hence we obtain quasilocal quantities for classical general relativity, 
too.

2. Notice that the topological condition for the existence of a
representation $\pi_{\Sigma ,\omega}$ is that both $i:\Sigma\looparrowright M$
and $\omega\in\Omega^2_c(M;\C )$ must represent non-trivial classes in 
$H_2(M;\Z )$ and $H^2(M;\C )$ respectively such that 
$\langle [\Sigma ],[\omega ]\rangle_M=
\frac{1}{2\pi\ii}\int_\Sigma\omega\not=0$. Hence in particular $\R^4$ 
or $S^4$ does not possess positive mass representations! However even if 
$[\Sigma_1]=[\Sigma_2]\in H_2(M;\Z )$ and 
$[\omega_1]=[\omega_2]\in H^2(M;\C )$ the resulting representations      
$\pi_{\Sigma_1 ,\omega_1}$ and $\pi_{\Sigma_2, \omega_2}$ are not 
unitarily equivalent in general. 
\end{remark}

\noindent Thirdly, apart from the tautological and positive mass 
{\it quantum} representations with unbroken symmetry ${\rm Diff}^+(M)$ 
there exist other ones what we call 
{\it classical representations} because in these representations the original 
vast symmetry group is spontaneously broken to a finite dimensional 
subgroup ${\rm Iso}^+(M,g)\subset{\rm Diff}^+(M)$ of an emergent metric 
$g$ on $M$.

\begin{theorem}

Take a perhaps non-compactly supported $\omega\in\Omega^2(M;\C )$ such 
that $\omega$ is non-degenerate along the whole $M$ moreover satisfies 
$\int_M\overline{\omega}\wedge\omega =1$. 

Then $\omega$ gives rise to a so-called {\rm classical representation} 
$\pi_\omega$ of $\fu (M)$ on a Hilbert space $\ch_\omega$ as follows:
\begin{itemize}

\item[(i)] $\ch_\omega$ also carries a unitary representation $U_\omega$ 
of the group $1\subseteqq{\rm Iso}^+(M,g)\subsetneqq{\rm Diff}^+(M)$ 
consisting of the isometries of the unitary metric $g$ on the complexified 
tangent bundle given by 
\[g (X,Y):=\frac{1}{2}\left(\omega (\overline{X},\ii Y)- 
\omega (\overline{\ii X}, Y)\right)\:\:\:\mbox{for all $X,Y\in C^\infty 
(M;TM\otimes_\R\C )$}\:\:\:.\] 
Moreover the state $\Omega :=[1]\in\ch_\omega$ corresponding to vanishing 
algebraic curvature $R=0$ satisfies 
$U_\omega (f)\Omega =\Omega$ for all $f\in {\rm Iso}^+(M,g)$; 

\item[(ii)] The distinguished splitting $\ch^+(M)\oplus\ch^-(M)$ via 
(anti)self-duality with respect to $g$ induces a splitting 
$\ch_\omega =\ch_\omega^+\oplus\ch_\omega^-$ into orthogonal subspaces obeyed 
by ${\rm Iso}^+(M,g)$. The distinguished quantum gravitational field $Q:=R_g$ 
in the sense of Definition \ref{kvantum} provided by the curvature of the 
metric $g$ acts on $\ch_\omega$. Moreover $\pi_\omega (R_g)$ obeys the 
splitting of $\ch_\omega$ if and only if $R_g$ does the same on 
$\ch^+(M)\oplus\ch^-(M)$ i.e., $R_g$ is a vacuum quantum gravitational field 
or in other words $g$ is a complexified Einstein metric on $M$. In particular 
if the metric $g$ is flat then $R_g=0$ also gives the invariant state 
$\Omega =[1]\in\ch_\omega$\:.
\end{itemize}

\label{klasszikus-reprezentacio}
\end{theorem}

\noindent{\it Proof.} (i) This time take another natural normalized linear 
functional $\Psi_\omega :\fu (M)\rightarrow\C$ by continuously extending a 
functional whose shape on elements 
$A\in{\rm End}(\Omega^2_c(M;\C ))\cap\fu (M)$ looks like 
\[A\longmapsto\Psi_\omega (A):=\int\limits_M\overline{\omega}\wedge 
(A\omega )=\langle\omega\:,\:A\omega\rangle_{L^2(M)}\]
provided by (\ref{integralas}). Exactly as in the proof of Theorem 
\ref{pozitiv-tomegu-reprezentacio} we can exploit the continuity of the 
functional to conclude from $\Psi_\omega (1^\divideontimes 1)=1$ that 
$\Psi_\omega$ is a positive functional on $\fu (M)$. 

Therefore applying again the GNS construction we come up with a 
reprsentation $\pi_\omega$ on a Hilbert space $\ch_\omega$. The metric 
also provides us with its isometry group 
$1\subseteqq{\rm Iso}^+(M,g)\subset{\rm Diff}^+(M)$. We construct a unitary 
representation $U_\omega$ of ${\rm Iso}^+(M,g)$ on $\ch_\omega$ as follows. 
First of all for any $f\in{\rm Iso}^+(M,g)$ we find 
$f^*\fu (M)(f^{-1})^*\subseteqq\fu (M)$. We define a representation on 
$\ch_\omega$ by $U_\omega (f)[A]:=[f^*A(f^{-1})^*]$. Moreover diffeomorphisms 
are unitary: $(f^* )^\divideontimes =(f^{-1})^*$ and in particular an 
isometry has the property $\omega =f^*\omega$ consequently 
\begin{eqnarray}
(U_\omega(f)[A],U_\omega (f)[B])_\omega&=&\int\limits_M\overline{\omega}
\wedge ((f^* A(f^{-1})^*)^\divideontimes (f^* B(f^{-1})^*\omega ))
=\int\limits_M\overline{\omega}\wedge (f^* A^\divideontimes B(f^{-1})^*\omega )
\nonumber\\
&=&\int\limits_M\overline{f^*\omega}\wedge (f^* A^\divideontimes B\omega )=
\int\limits_Mf^* (\overline{\omega}
\wedge (A^\divideontimes B\omega ))=\int\limits_M\overline{\omega}
\wedge (A^\divideontimes B\omega )\nonumber\\
&=&([A],[B])_\omega\nonumber
\end{eqnarray}
ensuring us that this representation is indeed unitary. 
$\Omega :=[1]\in\ch_\omega$ corresponding to the ``semiclassical 
gravitational vacuum'' $R=0$ is a (not necessarily unique) invariant vector.

(ii) Since $\fu (M)\subset{\rm End}(\ch^+(M)\oplus\ch^-(M))$ 
we get a decomposition of $\fu (M)$ as 
\[\fu (M)\cap\left({\rm End}(\ch^+(M))\oplus {\rm End}(\ch^-(M))\oplus
{\rm Hom}(\ch^+(M),\ch^-(M))\oplus {\rm Hom}(\ch^-(M),\ch^+(M))\right) .\]
Write an element $B\in{\rm End}(\Omega^2_c(M;\C ))\cap\fu (M)$ in the 
corresponding form as $B=\left(\begin{matrix}a&b\\
                   c&d
     \end{matrix}\right)$. It is easy to check that $\omega$ hence 
$\overline{\omega}$ is (anti)self-dual with respect to $g$ and the 
orientation on $M$ (on a complex manifold with its natural orientation 
$\omega$ is always self-dual, cf. \cite[Lemma 2.1.57]{don-kro}). Suppose now 
that $*_g\omega =\omega$. Then we obtain $B\omega =a\omega +c\omega $ with 
$a\omega\in\ch^+(M)$ as well as $c\omega\in\ch^-(M)$. Consequently exploiting 
the orthogonality of 
$\ch^+(M)$ and $\ch^-(M)$ we can expand $\Psi_\omega (B^\divideontimes B)$ 
and find
\[\Psi_\omega\left(\left(\begin{matrix} 
                           a^\divideontimes a+c^\divideontimes c& 
                           a^\divideontimes b+c^\divideontimes d\\
                           b^\divideontimes a+d^\divideontimes c&
                           b^\divideontimes b+d^\divideontimes d
                        \end{matrix}\right)\right) =
\Psi_\omega\left(\left(\begin{matrix}
                          a^\divideontimes a+c^\divideontimes c&0\\
                                             0&0
                          \end{matrix}\right)\right)\]
yielding that $\fu (M)\cap\left({\rm Hom}(\ch^-(M),\ch^+(M))
\oplus{\rm End}(\ch^-(M)\right)\subseteqq I_\omega$ where, as before, 
$I_\omega\subset\fu (M)$ is the Gelfand ideal of 
$\Psi_\omega$. Consequently $\ch_\omega$---being the completion of $\fu 
(M)/I_\omega$ with respect to the scalar product 
$(\:\cdot\:,\:\cdot\:)_\omega$---splits like 
$\ch_\omega^+\oplus\ch_\omega^-$ by completing 
$(\fu (M)\cap{\rm End}(\ch^+(M)))/I_\omega$ 
and $(\fu (M)\cap{\rm Hom}(\ch^+(M),\ch^-(M)))/I_\omega$ respectively. 
The two summands are orthogonal subspaces and the decomposition obviously 
satisfies $U_\omega (\ch^\pm_\omega )\subseteqq\ch^\pm_\omega$. The case of 
$*_g\omega =-\omega$ is similar. 

If $Q:=R_g$ is the curvature of $g$ regarded as a quantum gravitational 
field as in Definition \ref{kvantum} and $g$ is vacuum i.e., 
Einstein then we already know that $R_g(\ch^\pm (M))\subseteqq\ch^\pm (M)$. 
Moreover $R_g\in T_1\fu (M)$ acts on 
$\ch_\omega$ from the left by passing to the infinitesimal action of 
$\fu (M)$ on $\ch_\omega$ what we continue to denote by $\pi_\omega$. It 
then follows from 
$\left(\begin{matrix}
             p&0\\
             0&q\end{matrix}\right)
             \left(\begin{matrix}
              a&0\\ 
              c&0\end{matrix}\right)
           =\left(\begin{matrix}
              pa&0\\
              qc&0\end{matrix}\right)$ 
that for an Einstein metric $\pi_\omega (R_g)$ also satisfies $\pi_\omega 
(R_g)(\ch^\pm_\omega )\subseteqq\ch^\pm_\omega$. The particular case of the 
flat metric with $R_g=0$ gives the invariant state 
$\Omega =[1]\in\ch_\omega$ as well. $\Diamond$
\vspace{0.1in}

\noindent\begin{remark}\rm The usual axioms of algebraic quantum field 
theory (cf. e.g. \cite[pp. 58-60 or pp. 105-107]{haa}) typically make no 
sense in this very general 
setting. But for clarity we check them one-by-one in order to see in what 
extent our algebraic quantum field theory is more general than the usual 
ones.\footnote{We quote from Haag \cite[p. 60]{haa}: ``On the other hand the 
word $\gg$axiom$\ll$ suggests something fixed, unchangeable. This is 
certainly not intended here. Indeed, some of the assumptions are rather 
technical and should be replaced by some more natural ones as deeper insight 
is gained. We are concerned with a developing area of physics which is far from 
closed and should keep an open mind for modifications of assumptions, 
additional structural principles as well as information singling out a 
specific theory within the general frame.''}
 
\cite[Axiom {\bf A} on p. 106]{haa} can be translated to saying that the 
Hilbert space of a representation of the global generalized CCR algebra 
$\fu (M)$ also carries a unitary representation of the 
(spontaneously broken) space-time symmetry group of the theory which has 
been taken to be the whole diffeomorphism group here. We found 
three types of representations; here we discuss two of them. 

We constructed $\ch_{\Sigma ,\omega}$ carrying a {\it positive mass 
representation} $\pi_{\Sigma ,\omega}$ of $\fu (M)$ as well as a unitary 
representation $U_{\Sigma ,\omega}$ of the unbroken group ${\rm 
Diff}^+(M)$. However $\ch_{\Sigma ,\omega}$ does not possess a ${\rm 
Diff}^+(M)$-invariant state i.e., ``vacuum'' does not exist here. 
Nevertheless the Dougan--Mason quasilocal translations of 
$i:\Sigma\looparrowright M$ give rise to quasilocal energy-momentum 
$4$-vectors $P_{C,\omega ,A}$ in a manner that the state corresponding to 
the classical gravitational vacuum has vanishing energy-momentum as one 
expects. This is interesting because the concepts of mass and energy are 
quite problematic in classical general relativity as well as that of the 
vacuum in general quantum field theories. But recall that this 
construction---which mixes ideas of quasilocal constructions in classical 
general relativity \cite{dou-mas, sza} and standard GNS representation 
theory of $C^*$-algebras---contains a technical ambiguity namely a choice 
of a complex structure on an immersed surface in $M$. However one expects 
the whole machinery to be independent of this choice. We treat this 
problem in Sect. \ref{four}.

We also constructed $\ch_\omega$ carrying a {\it classical representation} 
$\pi_\omega$ of $\fu (M)$ together with a unitary representation $U_\omega$ of 
the spontaneously broken group ${\rm Iso}^+(M,g)\subset{\rm Diff}^+(M)$ 
provided by an emergent metric $g$ on $M$. This representation gives back the 
classical picture. It also possesses a (probably not unique) invariant state 
$\Omega\in\ch_\omega$ but this time we lack the concept of energy-momentum 
hence we cannot call this state as the ``vacuum''. 

\cite[Axioms {\bf B} and {\bf C} on p. 107]{haa} dealing with the additivity of
local algebras and their hermiticity by construction hold here. 

\cite[Axiom {\bf D} on p. 107]{haa} can be translated to saying that 
since the diffeomorphism group is the symmetry group of the theory, it 
acts on the net of local algebras like  
\begin{equation}
f^*\fu (U)(f^{-1})^* =\fu (f(U))
\label{kovariancia}
\end{equation}
for all $f\in{\rm Diff}^+(M)$ i.e., symmetry transformations map the 
local algebra of a region to that one of the transformed region. This 
continues to be valid here.

\cite[Axiom {\bf E} on p. 107]{haa} holds in a trivial way as an unavoidable 
consequence of the vast diffeomorphism invariance. It is easy to see 
that $[\fu (U),\fu (V)]=0$ if and only if $U\cap V=\emptyset$. Indeed, 
demanding (\ref{kovariancia}) to be valid we can see that regardless 
what $\fu (U)$ actually is, it must commute with diffeomorphisms being the 
identity on $U$; consequently if $A\in\fu (U)\subset\cb (M)$ then 
$A\vert_{\Omega^2_c(M\setminus U;\C )}\in {\mathfrak Z}(\cb (M\setminus U))=
\C\:{\rm Id}_{\Omega^2_c(M\setminus U;\C )}$. But 
$\Omega^2_c(V;\C )\subset\Omega^2_c (M\setminus U;\C )$ if 
$U\cap V=\emptyset$ hence the assertion follows. Therefore there is no 
causality hence no dynamics present here. Hence the reason we prefer to use 
Riemannian metrics over Lorentzian ones throughout the paper (although 
emphasize again that all conclusions hold for Lorentzian metrics as well). 
We can also physically say that this theory represents a very elementary 
level of physical reality where even no causality exists yet. Causality 
should emerge through breaking of the diffeomorphism symmetry. This symmetry 
breaking has been carried out in the case of the classical representations. 

\cite[Axiom {\bf F} on p. 107]{haa} This completeness requirement claims for 
the validity of Schur's lemma i.e., in a representation the only bounded 
operator which commutes with all quantum observables should be a 
multiple of the identity operator. This holds if the representation of 
$\fu (M)$ in question is irreducible. 

\cite[Axiom {\bf G} on p. 107]{haa} about ``primitive causality'' has no 
meaning in this general setting.

\end{remark}


\section{Positive mass representations and conformal field theory}
\label{four}


Theorem \ref{pozitiv-tomegu-reprezentacio} allows us to make a link 
with conformal field theory. We obtained representations 
$\pi_{\Sigma ,\omega}$ of the algebra of global observables $\fu (M)$ 
constructed by standard means from a smooth immersion 
$(\Sigma ,p_1,\dots ,p_n)$ of a surface $\Sigma$ into $M$ and a regular 
element $\omega\in\Omega^2_c(M;\C )$. If a complex structure 
$C=C(\Sigma )$ is put onto the surface as well then the quasilocal 
energy-momentum $P_{C,\omega ,A}\in H^0(C;\co (E))$ and mass 
$m_{C,\omega ,A}\in\R^+\cup\{ 0\}$ of a non-zero state 
$[A]\in\ch_{\Sigma ,\omega}$ can be defined enriching $\pi_{\Sigma ,\omega}$ 
further to a positive mass representation. However 
on physical grounds we expect the whole construction to be independent of 
these technicalities i.e., any choice of these complex structures have to 
result in the same construction. Following Witten \cite{wit} this means 
that a conformal field theory lurks behind the curtain. We can indeed find 
this theory which however turns out to be a very simple topological 
conformal field theory in the sense that its Hilbert space is finite 
dimensional and the correlation functions are insensitive for the 
insertion of marked points i.e., how the immersion looks like.

In constructing this topological conformal field theory we will follow G. 
Segal \cite{seg}. That is first construct a ``modular functor extended 
with an Abelian category possessing a symmetric object'' (cf. 
\cite[Definition 5.1.12]{bak-kir}) in particular and 
\cite[Chapters 5 and 6]{bak-kir} in general). In other words we have to 
construct an assignment 
\begin{equation} 
\tau : (\Sigma ,p_1,\dots,p_n)\longmapsto \tau (\Sigma ,p_1,\dots ,p_n) 
\label{cft} 
\end{equation} 
which somehow associates to 
surfaces with marked points finite dimensional complex vector spaces 
satisfying certain axioms. Consider a positive mass representation 
$\pi_{\Sigma ,\omega}$ of $\fu (M)$ constructed out of 
$(\Sigma ,p_1,\dots,p_n,\omega )$ as in Theorem 
\ref{pozitiv-tomegu-reprezentacio}. Recall that the marked points 
$p_i\in\Sigma$ correspond the multiple points of the immersion 
$i:\Sigma\looparrowright M$ (the case $(\Sigma ,\emptyset ,\omega)$ is an 
embedding). Then to a positive mass representation 
of $\fu (M)$ a holomorphic vector bundle $\ce$ of spaces of conformal blocks 
$\tau (\Sigma ,p_1,\dots,p_n)$ over the coarse moduli space 
$\cm_{g,n}$ of complex structures on $(\Sigma ,p_1,\dots,p_n)$ will 
be assigned in manner that if $0\not= [A]\in\ch_{\Sigma ,\omega}$ is a state 
then its quasilocal energy-momentum $4$-vector $P_{C,\omega ,A}$ gives 
rise to a section $P_{\Sigma ,\omega ,A}$ of $\ce$. This section will be 
moreover (projectively) flat with respect to the natural (projectively) flat 
connection $\nabla$ on $\ce$ (the Knizhnik--Zamolodchikov connection). 
In other words the quasilocal energy-momentum $4$-vector 
gives rise to a conformal block in this conformal field theory.

We begin with the following simple observation (an elementary version of 
Uhlenbeck's singularity removal theorem \cite{uhl}).

\begin{lemma}
Take any compact Riemann surface $C=C(\Sigma )$ with distinct marked points 
$p_1,\dots,p_n\in C$ and a holomorphic unitary vector bundle $F'$ over 
$C\setminus\{ p_1,\dots,p_n\}$. Let $s'\in H^0(C\setminus\{ 
p_1,\dots,p_n\};\co (F'))$ be a meromorphic section with the property 
$\Vert s'\Vert_{L^2_{loc}(C)}<+\infty$ i.e., having locally finite energy 
over $C$. 

If $s'$ is singular in $p_i\in C$ then one can find a local gauge 
transformation about this point such that the gauge transformed section 
extends holomorphically across it i.e., pointlike singularities of locally 
finite energy meromorphic sections over $C$ are removable. More precisely 
there exists a unique unitary holomorphic vector bundle $F$ over $C$ 
satisfying $F\vert_{C\setminus\{ p_1,\dots,p_n\}}\cong F'$ so that for any 
locally finite energy section $s'\in H^0(C\setminus\{ p_1,\dots,p_n\};
\co (F'))$ there exists a section $s\in H^0(C;\co (F))$ with the property 
$s\vert_{C\setminus\{ p_1,\dots,p_n\}}$ is gauge equivalent to $s'$. 
\label{uhlenbeck}
\end{lemma}

\noindent {\it Proof.} First we prove the existence of the unique 
extendibility of the unitary bundle $(F',h')$. Consider a local holomorphic 
coordinate system $(U_i,z)$ on $C$ such that $z(U_i)=D(0)\subset\C$ some open 
disc about the origin and $U_i$ contains only one marked point $p_i\in U_i$ 
satisfying $z(p_i)=0$. Cutting out the open neighbourhood $U_i\subset C$ 
of $p_i$ we obtain a manifold-with-boundary $C\setminus U_i$ and 
$\partial (C\setminus U_i)\cong S^1$. Consider the restriction 
$(F',h')\vert_{\partial (C\setminus U_i)}$ regarded as a smooth 
${\rm U}(k)$-bundle over $S^1$. Taking a smooth local trivialization the 
corresponding smooth local transition function of $(F',h')\vert_{\partial 
(C\setminus U_i)}$ gives rise to a monodromy map 
$\mu_i :S^1\rightarrow{\rm U}(k)$ where $k={\rm rk}\:F'$. However 
$\pi_0 ({\rm U}(k))\cong 1$ hence this monodromy map together with 
its derivatives along $S^1$ extends over $p_i$ as the identity consequently 
$(F',h')\vert_{U_i\setminus\{ p_i\}}$ can be extended over this point as a 
smooth unitary vector bundle $(F_i ,h_i)\vert_{U_i}$. Consider a smooth 
trivialization $F_i\vert_{U_i}\cong U_i\times\C^k$ and write in this
smooth gauge the restriction of the partial connection defining the 
holomorphic structure on $F'$ as 
$\overline{\partial}_{F'}\vert_{U_i\setminus\{ p_i\}}=
\overline{\partial}+\alpha'_{U_i\setminus\{ p_i\}}$. Then the Hermitian 
scalar product on $F_i$ satisfies 
\[\overline{\partial}_{F_i}(h_i\vert_{U_i\setminus\{ p_i\}})=
\overline{\partial}(h_i\vert_{U_i\setminus\{ p_i\}})+
\alpha'_{U_i\setminus\{ p_i\}}(h_i\vert_{U_i\setminus\{ p_i\}})=0\]
and $h_i\vert_{U_i\setminus\{p_i\}}$ extends smoothly over $p_i$ as 
$h_i\vert_{U_i}$. Therefore 
$\alpha_{U_i}:=-(\overline{\partial}h_i\vert_{U_i})(h_i\vert_{U_i})^{-1}$ on 
$U_i$ defines a smooth extension of $\alpha '_{U_i\setminus\{ p_i\}}$ 
over $p_i$ in a manner that $\overline{\partial}_{F_i}\vert_{U_i}:=
\overline{\partial}+\alpha_{U_i}$ is the restriction of a compatible partial 
connection $\overline{\partial}_{F_i}$ yielding a compatible holomorphic 
structure on $(F_i, h_i)$. Performing this procedure around every marked 
points we obtain a unique unitary holomorphic vector bundle i.e., 
$(F,h ,\overline{\partial}_F)$ with $\overline{\partial}_Fh=0$.

Now we come to the extendibility of sections. Compatibility provides us 
that in a local holomorphic trivialization $F\vert_U\cong U\times\C^k$ the 
coefficients of $h\vert_U$ are holomorphic functions. Performing a 
${\rm GL}(k,\C )$-valued holomorphic gauge transformation if necessary we can 
pass to a local holomorphic trivialization in which $h\vert_U$ has the 
standard form. Take any holomorphic section of $F$ or equivalently, a 
meromorphic section of $F$ with singularities in the marked points i.e., 
pick any 
\[s'\in H^0(C\setminus\{ p_1,\dots,p_n\};\co (F'))\cong 
H^0(C\setminus\{p_1,\dots,p_n\};\co (F))\] 
with local shape $s'\vert_U(z)=s'^1(z)f_1+\dots +s'^k(z)f_k$ in this local 
trivialization. Since $s'\vert_U$ is holomorphic outside $0\in\C$ each 
components $s'^j:U\rightarrow\C$ admit Laurent expansions
\[s'^j(z)=\sum\limits_{m=-N^j}^{+\infty}a^j_mz^m,\:\:\:\:\:a^j_m\in\C\:\:\:.\]
Moreover the local $L^2$-norm of the section in this special gauge looks like
\[\Vert s'\Vert^2_{L^2(U)}=\frac{1}{2\pi\ii}
\int\limits_U\left( \vert s'^1(z)\vert^2+\dots
+\vert s'^k(z)\vert^2\right)\omega\vert_U=
\int\limits_U\left( \vert s'^1(z)\vert^2+\dots 
+\vert s'^k(z)\vert^2\right)\varphi_U(z,\overline{z})
\dd z\wedge\dd\overline{z}\]
where $\varphi_U$ is a smooth nowhere vanishing function on $U$. 
Assume that the section has locally finite energy. On substituting the 
above expansions into this integral the finiteness then dictates to conclude 
that $N^j=0$ for all $j=1,\dots,k$ and $i=1,\dots,n$ hence in fact $s'$ is 
holomorphic over the whole $C$ as desired. $\Diamond$
\vspace{0.1in}

\noindent Now we turn to the construction of the relevant modular functor. 
Suppose that $\Sigma\looparrowright M$ is a compact surface without 
boundary. Choose any complex structure $C=C(\Sigma )$ on it and $n$ distinct 
marked points $p_1,\dots ,p_n\in C$ given by multiple-points of the immersion. 
Let $E':=TM\otimes_\R\C\vert_{C\setminus\{p_1,\dots,p_n\}}$ be a 
holomorphic unitary bundle over the punctured surface. Or rather more 
generally, if $C=\sqcup_iC_i$ is an abstract compact non-punctured Riemann 
surface as in Theorem \ref{pozitiv-tomegu-reprezentacio} with 
connected components $C_i$ then let $E$ be a holomorphic unitary 
bundle over $C$ with ${\rm rk}_\C(E\vert_{C_i})=4$, 
${\rm deg}(E\vert_{C_i})=0$ and $\dim_\C H^0(C;\co (E))=4$. Then in terms of 
the restricted bundle $E':=E\vert_{C\setminus\{p_1,\dots,p_n\}}$ our 
choice is as follows: 
\begin{equation}
\tau (\Sigma ,p_1,\dots,p_n):= 
\left\{\begin{array}{ll}
{\rm Cliff}\left( H^0\left( C\setminus\{p_1,\dots,p_n\}; 
\co (E')\right)\cap L^2_{loc}(C;E)\right)& 
   \!\mbox{if $(\Sigma ,p_1,\dots,p_n)\not=\emptyset$\:\:\:;} \\
             & \\ \C & \!\mbox{if $(\Sigma ,p_1,\dots,p_n)=\emptyset$} 
\end{array}\right.
\label{konformal.blokk} 
\end{equation} 
that is, this vector space is the underlying vector space of the complex 
Clifford algebra of the scalar product space 
\[\left( H^0(C\setminus\{p_1,\dots,p_n\};\co (E'))\cap 
L^2_{loc}(C;E)\:,\:(\:\cdot\:,\:\cdot\:)_{L^2(C)}\right)
\cong\C^4_{{\rm Hermite}}\]
consisting of vector fields on $M$ which, upon restriction to $C$, are 
holomorphic except in the marked points and have locally finite energy.

\begin{lemma} Let $(\Sigma ,p_1,\dots,p_n)$ be a smooth surface with 
marked points and take a complex structure $C=C(\Sigma )$ rendering it 
a Riemann surface with marked points $(C,p_1,\dots,p_n)$. Also take the 
holomorphic unitary bundle $E'$ over $C\setminus\{p_1,\dots,p_n\}$ 
as before. Attach to every marked point $p_i\in C$ the single label 
\[\nu :=\{\mbox{a holomorphic section of $E'$ has a finite energy 
singularity in $p_i\in C$}\}\:\:\:.\] 
Then the assignment (\ref{cft}) with the choice (\ref{konformal.blokk}) is a 
modular functor which is not normalized in the sense that $\tau 
(S^2,\emptyset )={\rm Cliff}(H^0(\C P^1;\co (E')))$ instead of $\tau 
(S^2,\emptyset )=\C$. 

Moreover the vector spaces $\tau (\Sigma ,p_1,\dots,p_n)$ fit together into 
a trivial holomorphic vector bundle $\ce$ over the coarse moduli space 
$\cm_{g,n}$ of genus $g$ Riemann surfaces with $n$ marked points 
carrying a flat connection $\nabla$ (the Knizhnik--Zamolodchikov connection). 
The vector $P_{C,\omega ,A}\in H^0(C;\co (E))$ is the value at 
$C\in\cm_{g,n}$ of a section $P_{\Sigma ,\omega ,A}$ of this 
bundle over $\cm_{g,n}$ satisfying $\nabla P_{\Sigma ,\omega ,A}=0$.
\end{lemma}

\noindent{\it Proof.} We check the three relevant axioms of 
\cite[Definition 5.1.2]{bak-kir}. First of all Lemma \ref{uhlenbeck} 
yields that if $(\Sigma ,p_1,\dots ,p_n)\not=\emptyset$ then 
\[\tau (\Sigma ,p_1\dots,p_n)\cong{\rm Cliff}(H^0(C;\co (E)))\] 
consequently the vector spaces are finite dimensional. It also readily 
follows from (\ref{konformal.blokk}) that
\[\tau ( (\Sigma_1, p_1,\dots,p_n)\sqcup 
(\Sigma_2,q_1,\dots,q_m))\cong \tau (\Sigma_1, 
p_1,\dots,p_n)\otimes_\C\tau (\Sigma_2,q_1,\dots,q_m)\] 
as vector spaces, in agreement with \cite[ part (iii) of Definition 
5.1.2]{bak-kir}. The second axiom to check is \cite[ part (iv) of 
Definition 5.1.2]{bak-kir} which is the glueing axiom. Let $\gamma\subset 
(\Sigma ,p_1,\dots,p_n)$ be a closed curve without self-intersections. Cut 
$(\Sigma ,p_1,\dots,p_n)$ along $\gamma$. The resulting surface has naturally 
the structure of a not necessarily connected punctured surface 
$(\tilde{\Sigma} ,p_1,\dots,p_n, q_1,q_2)$ where the two 
new marked points $q_1, q_2$ come from the circle $\gamma$. Putting 
$\tilde{E}:=E\vert_{C\setminus (\{p_1,\dots,p_n\}\cup\gamma )}$ into 
(\ref{konformal.blokk}) by the aid of Lemma \ref{uhlenbeck} we obtain 
that locally finite energy meromorphic sections on $(\tilde{C} 
,p_1,\dots,p_n, q_1,q_2)$ correspond to those on $(C, p_1,\dots,p_n)$ 
consequently, taking into account that there is only a single label $\nu$ 
with its meaning,
\[\tau (\tilde{\Sigma},p_1,\dots,p_n, q_1,q_2)\cong \tau 
(\Sigma ,p_1,\dots,p_n)\] 
hence the glueing axiom holds in a trivial way here.

The third axiom to check is the functorial behaviour under 
diffeomorphisms \cite[ part (ii) of Definition 5.1.2]{bak-kir}. In turn 
this is equivalent to checking the existence of a 
Knizhnik--Zamolodchikov connection. Let $\cm_{g,n}$ be the 
coarse moduli space of connected non-singular Riemann surfaces of genus 
$g$ and $n$ marked points. We take a complex vector bundle $\ce$ over 
$\cm_{g,n}$ whose fibers over $(C,p_1,\dots, p_n)\in\cm_{g,n}$ are the 
individual spaces of conformal blocks 
$\tau (\Sigma ,p_1,\dots,p_n)$ constructed from the holomorphic bundle 
$E'$ over $C\setminus\{p_1,\dots, p_n\}$ or equivalently $E$ over $C$. Recall 
that $M$ is acted upon by its diffeomorphism group. Hence the subgroup 
${\rm Diff}^+_\Sigma (M)\subset{\rm Diff}^+(M)$ consisting of 
$\Sigma$-preserving diffeomorphisms acts on the real smooth 
punctured surface such that it deforms its complex structure i.e., 
$(\Sigma ,p_1,\dots,p_n)$ and $(f(\Sigma ) ,f(p_1),\dots,f(p_n))$ correspond in 
general to different points in $\cm_{g,n}$. This subgroup also 
acts on $C^\infty (C; E)$ by pullback. Consequently it 
transforms the subspaces $\tau (\Sigma ,p_1,\dots ,p_n )
\cong{\rm Cliff}(H^0(C;\co (E)))\subset{\rm Cliff}(C^\infty (\Sigma ;E))$ 
giving rise to linear isomorphisms 
\[\tau (\Sigma ,p_1,\dots ,p_n )
\cong\tau (f(\Sigma ) ,f(p_1),\dots ,f(p_n )) \:\:\:\:\:\mbox{for all 
$f\in {\rm Diff}^+_\Sigma (M)$}\:\:\:.\] 
These linear isomorphisms can be interpreted as parallel translations along 
$\ce$ by a flat connection $\nabla$ called the Knizhnik--Zamolodchikov 
connection. Note that since the representation of $\Sigma$-preserving 
diffeomorphisms on $C^\infty (\Sigma ;E)$ is not only 
projective but in fact a true representation, the resulting connection is not 
only projectively but truely flat on $\ce$. In particular the bundle $\ce$ as 
a complex vector bundle is trivial over $\cm_{g,n}$ but is 
equipped with a holomorphic structure. Via Lemma \ref{uhlenbeck} the 
holomorhic section $P_{C,\omega, A}\in H^0(C;\co (E))$ can be regarded as 
a meromorphic one i.e., $P_{C,\omega, A}\in 
H^0(C\setminus\{ p_1,\dots,p_n\};\co (E'))$. 
Define a section $P_{\Sigma ,\omega ,A}$ of $\ce$ on $\cm_{g,n}$ by 
putting $P_{\Sigma ,\omega ,A}(C):=P_{C,\omega ,A}$. It follows from the 
invariance of the definition (\ref{dougan--mason}) of the quasilocal 
energy-momentum $4$-vector 
\begin{eqnarray}
P_{C,\omega ,A}&\in &H^0(C;\co (E))\subset
{\rm Cliff}\left (H^0(C;\co (E))\right)\cong\nonumber\\
&&{\rm Cliff}\left( H^0\left( C\setminus\{p_1,\dots,p_n\};   
\co (E')\right)\cap L^2_{loc}(C;E)\right)
=\tau(\Sigma ,p_1,\dots,p_n)\nonumber
\end{eqnarray}
under diffeomorphisms that as the complex structure varies 
$P_{\Sigma ,\omega ,A}$ of $\ce$ satisfies $\nabla P_{\Sigma ,\omega ,A}=0$ 
i.e., is parallel for the Knizhnik--Zamolodchikov connection. 

We conclude that the assignment (\ref{cft}) with 
(\ref{konformal.blokk}) is a $C$-extended modular functor as in 
\cite[Definition 5.1.2]{bak-kir} i.e., a weakly 
conformal field theory {\it \'a la} G. Segal \cite{seg}. $\Diamond$
\vspace{0.1in}

\noindent After having constructed the modular functor, we find the 
vector space on which it acts hence exhibit the conformal field theory given 
by (\ref{cft}) and (\ref{konformal.blokk}). This step is very simple: 
the space $(\Sigma ,p_1,\dots ,p_n)$ identified with an oriented smooth 
cobordism between the disconnected compact oriented 
$1$-manifolds $S^1_{p_1}\sqcup\dots\sqcup S^1_{p_k}$ and 
$S^1_{p_{k+1}}\sqcup\dots\sqcup S^1_{p_n}$. To the oriented $1$-manifold 
$S^1_{p_1}\sqcup\dots\sqcup S^1_{p_k}\sqcup 
(S^1_{p_{k+1}})^*\sqcup\dots\sqcup (S^1_{p_n})^*$, regardless what it 
actually is, we associate the finite dimensional complex vector space 
$S\otimes_\C S^*$ where $S$ is the unique irreducible complex 
Clifford-module of $\tau (\Sigma ,p_1,\dots,p_n)$. The resulting conformal 
field theory is a topological one because its state space is finite 
dimensional and its correlation functions are insensitive for the insertion 
of marked points (due to Lemma \ref{uhlenbeck}).

\end{document}